%% file: RingAllreduce-arxiv-v1.tex
\documentclass[12pt,onecolumn]{IEEEtran} %
\usepackage{cite}
\usepackage{amsmath,amssymb,amsfonts}
\usepackage{graphicx}
\usepackage{textcomp}
\usepackage{xcolor}

\usepackage{bm}
\usepackage{subfig}
\usepackage{color, colortbl}
\usepackage{url}
\usepackage{amsmath}
\usepackage{amsthm}
\usepackage{amssymb}

\usepackage{tikz} 
\usepackage{pgfplots}
\usepackage[normalem]{ulem}
\pgfplotsset{compat=1.5}

\usepackage{graphics}

\usetikzlibrary{arrows}

\newtheorem{lemma}{Lemma}
\newtheorem{corollary}{Corollary}

\newtheorem{proposition}{Proposition}

\begin{document}

\title{Performance Analysis of Distributed Training with Ring AllReduce}
\author{Fahimeh Fazel, Almut Burchard, J\"{o}rg Liebeherr\thanks{
F. Fazel (fahimeh.fazel@mail.utoronto.ca) and J. Liebeherr (jorg@ece.utoronto.ca)  are with the Department of Electrical and Computer Engineering, University of Toronto. A. Burchard (almut@math.toronto.edu) is with the Department of Mathematics,  University of Toronto.}
}


\maketitle

\begin{abstract}
Distributed machine learning reduces training time by spreading computations across many workers. Recent measurement studies of distributed training have reported that synchronization and communication overheads leave GPUs substantially underutilized. In this paper, we study the randomness arising from variations in computation, communication, and synchronization as a root cause of worker idle times in distributed training. Our analysis considers Ring AllReduce, a widely used bandwidth-optimal collective communication algorithm. We represent Ring AllReduce as a Stochastic Petri Net (SPN) with exponentially distributed delays. The SPN induces a continuous-time (absorbing) Markov chain with a state space that grows exponentially in the number of workers. Observing that the fraction of idle workers quickly settles to a stable value, we compute the long-run probability that a worker is idle without enumerating the state space. For homogeneous workers, we show that the idle probability of workers converges to 50\%. We extend our analysis to settings with less randomness, where delays have both deterministic and random components, and to systems with stragglers. To characterize the fluctuations of completion times beyond their mean, we couple our SPN model to a special case of the totally asymmetric simple exclusion process (TASEP). Exact asymptotic results for TASEP yield the distribution of completion times, showing that performance metrics of Ring AllReduce scale differently than suggested by the central limit theorem.
\end{abstract}

\input{1_Introduction_Nov23}
\input{2_Background_and_Related_Work_Nov23}

\input{3_Modeling_Serverless_Distributed_Training_Using_SPN_Nov23}

\input{4_Analysis_of_Idle_Times_Nov23}

\input{5A_Mapping_to_TASEP_Nov25}
\input{6_Conclusion}

\section*{Acknowledgments}

This work has been supported in part by two NSERC Discovery grants and 
the NSERC Centre of Excellence for Networking Innovation in Toronto (CENIT).
The authors wish to thank Jeremy Quastel 
for helpful discussions about TASEP. 

\vspace{1cm}
\bibliographystyle{IEEEtran}
\bibliography{IEEEabrv,bibliography}

\end{document}

%% file: 1_introduction_Nov23.tex
\section{Introduction}
\label{sec:introduction}

The rapid growth in the size and computational demands of machine 
learning models has made distributed training increasingly important. 
By distributing the computation of a training job across multiple graphical processing units (GPUs) or 
compute nodes, so-called {\it workers}, distributed machine learning (DML) 
reduces the time for training a model and supports models whose 
resource requirements exceed the capacity of a single worker. 
At the same time, DML introduces overheads 
due to the communication required to exchange and synchronize intermediate 
results among workers. As the number and computational power of workers increase, communication and synchronization 
accounts for a significant fraction of training time. 

Measurement studies of DML workloads report substantial underutilization and idle
time of GPU resources~\cite{jeon2019analysis,hu2024characterization}. For example, Jeon et al.~\cite{jeon2019analysis} analyze a two-month trace of a
multi-tenant GPU cluster and report that the average hardware
utilization of GPUs is around $52\%$. This is attributed in large parts 
to  synchronization during model training, where GPUs remain
idle while waiting for other GPUs. 
Hu et al.~\cite{hu2024characterization} analyze a six-month trace of an LLM development datacenter and find that approximately $30\%$ of GPUs are in an idle
state~\cite{hu2024characterization}.
Some studies use Model FLOPs Utilization~(MFU), which measures the fraction of  peak floating-point performance, as efficiency metric for GPUs. 
Jian et al.~\cite{jiang2024megascale}  reports an MFU of $55.2\%$ when training a 
175B-parameter large language model (LLM) on
$12{,}288$~GPUs.  A study from Meta~\cite{Grattafiori2024} reports an overall
MFU of $38\%$ to $43\%$ for training Llama~3.1 405B across a 
training network with up to $16{,}384$ H100 GPUs. 
These observations raise a central question: why do GPUs remain underutilized even in optimized distributed training systems? 

The reasons for low resource utilization in distributed training 
are manifold, and ultimately stem from transient or persistent asymmetries in computation, communication, and synchronization across workers. In this paper, we explore   
 randomness resulting from the collection of such asymmetries, as the root cause of 
 worker idle time. Specifically, we present a stochastic analysis of  the impact of randomness on the idle times of workers for the {\it Ring AllReduce} algorithm. 

In distributed training, workers exchange gradient tensors to compute a consistent update to model parameters.
In practice,  gradient tensors are summed or averaged, 
and the result is disseminated to all workers. 
This operation is referred to as {\it AllReduce}. 
The semantics of AllReduce  can be realized in a centralized or decentralized fashion. 
In a centralized system, workers send gradients
to a parameter server, which aggregates the gradients
and returns the results to 
the workers. 
The drawback of a centralized design with a parameter server 
is that it limits scalability and may cause network congestion
~\cite{AlqahtaniDemirbas2019}.
A decentralized design avoids the bottleneck of a parameter server by enabling
workers to exchange data directly through coordinated collective communication.
Ring AllReduce~\cite{Patarasuk2009,ringallreduce-uber} is a widely used decentralized algorithm, where  workers are arranged in a logical ring and exchange data only with their 
neighbors in the ring. Ring AllReduce is bandwidth-optimal in the sense that it matches a lower bound for the data volume that must be transmitted by each worker~\cite{Patarasuk2009}.
If the logical ring is mapped onto a network without creating a bottleneck, a 
deterministic 
Ring AllReduce operation can yield a full utilization of workers~\cite{Patarasuk2009}. 

In this paper we develop stochastic analytical models of Ring AllReduce that
characterize idle times of workers due to dependencies created by the ring structure. 
Whereas previous analyses of Ring AllReduce often introduce simplifying assumptions, such as global synchronization barriers and/or independence between worker completion times, 
the main contribution of our analysis is that we capture the coupling of operations throughout all steps of 
the algorithm. 
We first present a Stochastic
Petri Net (SPN) representation that captures the
dependency structure among workers under exponentially 
distributed delays. 
The SPN model induces a continuous-time Markov chain (CTMC) 
with a state space that grows exponentially with the number of workers, making
a direct solution prohibitive even for a modestly sized ring. 
A simulation of the Ring AllReduce model shows that the fraction of idle 
workers quickly reaches a stable state. 
This, in turn, can be exploited by conducting a steady-state analysis: 
We are able to compute the steady-state probability that a worker is idle directly without the need to explore the entire state space. 
An important finding is that, with homogeneous workers, the idle probability of 
a worker during Ring AllReduce converges to 50\% as the number of workers 
is increased. 
Using the idle probability, we also obtain a reasonably accurate 
estimate for the average completion time of  Ring AllReduce. 
We are able to extend our analysis to a system with less randomness, where 
delays have a deterministic and a random component. 
We also analyze the impact of the presence of slower workers, so-called {\it stragglers}, 
on the Ring AllReduce algorithm. 

To capture the statistical fluctuations of the completion times of Ring AllReduce (as opposed to their mean value) we develop a novel analysis where we couple the SPN to a special case of 
the totally asymmetric exclusion process (TASEP)~\cite{spitzer1970}, a stochastic process used to model interacting particle systems in mathematical physics and 
other areas. Recent results in probability theory have yielded 
exact asymptotic results for TASEP~\cite{quastel2007}, which help us to precisely estimate the distribution of completion times 
of Ring AllReduce. TASEP theory 
also explains why the performance metrics of Ring AllReduce do not scale in 
accordance with the central limit theorem.

The remainder of the paper is organized as follows.
In Sec.~\ref{sec:related_work}, we review the Ring AllReduce algorithm and discuss previous  modeling efforts.
In Sec.~\ref{sec:spn-model}, we develop the SPN model and induced CTMC and 
derive invariances of the model. 
In Sec.~\ref{sec:idle_time_analysis}, we present our analysis of the idle 
probability and average completion times.  
In Sec.~\ref{sec:tasep}, we 
provide a mapping of the SPN model to a TASEP, and obtain the distribution of 
completion times in Ring AllReduce. 
We summarize our findings in Sec.~\ref{sec:conclusion} and present our conclusions. 

%% file: 2_Background_and_Related_Work_Nov23.tex
\section{Background and Related Work}
\label{sec:related_work}

Ring AllReduce is a serverless algorithm 
for implementing the AllReduce collective communication primitive. 
It has its origin in high-performance computing~\cite{Patarasuk2009} and is now widely used in distributed machine learning, where it replaces reductions with a centralized parameter server~\cite{ringallreduce-baidu,ringallreduce-uber}. 
Ring-AllReduce is bandwidth-optimal in the sense that, with $N$ nodes, each node sends and receives  $2(N-1)/N$ times the amount of data being reduced,  
so the communication overhead of each worker remains nearly constant as the number of workers is increased.  
In the following we will refer to the execution of the 
Ring Allreduce algorithm as `Ring AllReduce operation.'

\begin{figure}[t!]
\centering 
\subfloat[]{\includegraphics[width=0.24\textwidth]{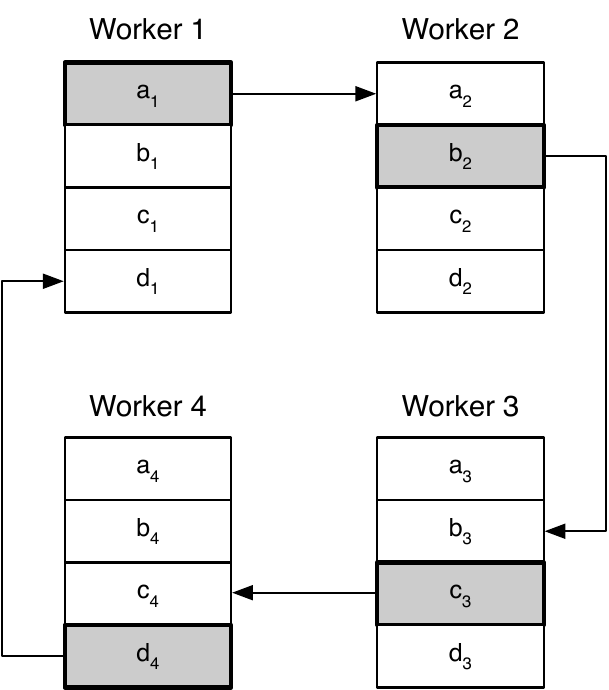}
\label{fig:Ringallreduce-1}}
\subfloat[]{\includegraphics[width=0.24\textwidth]{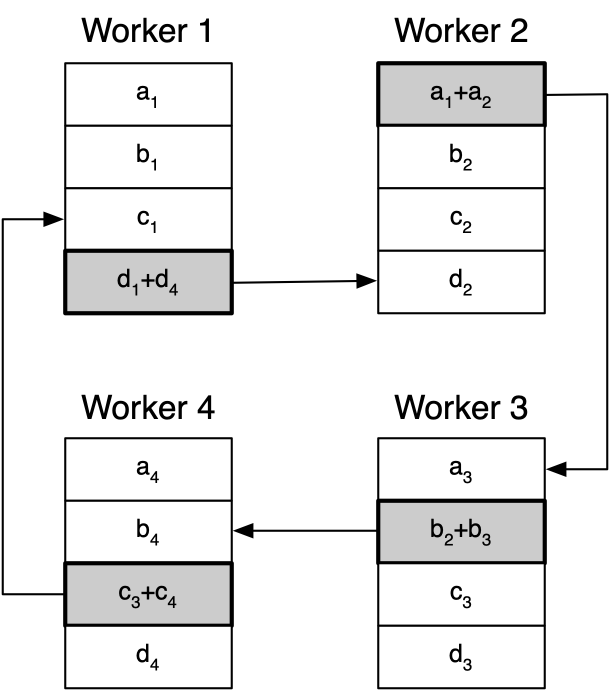}
\label{fig:Ringallreduce-2}}
%
%
%
%
\subfloat[]{\includegraphics[width=0.24\textwidth]{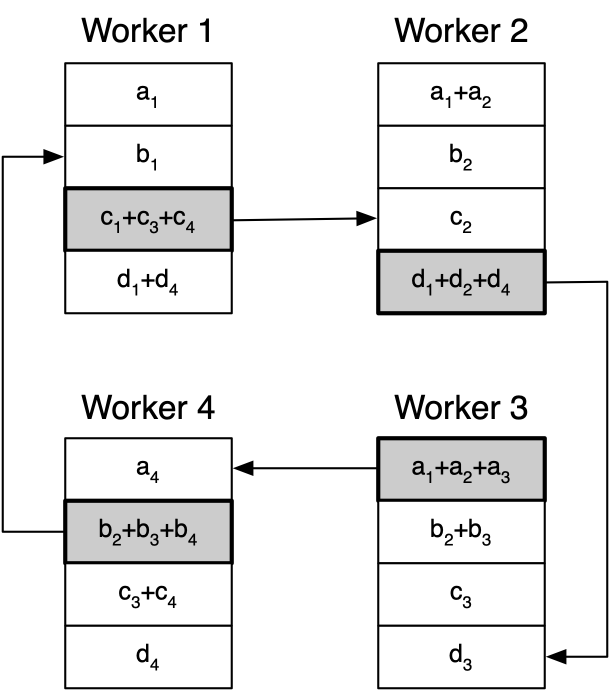}
\label{fig:Ringallreduce-3}}
\subfloat[]{\includegraphics[width=0.24\textwidth]{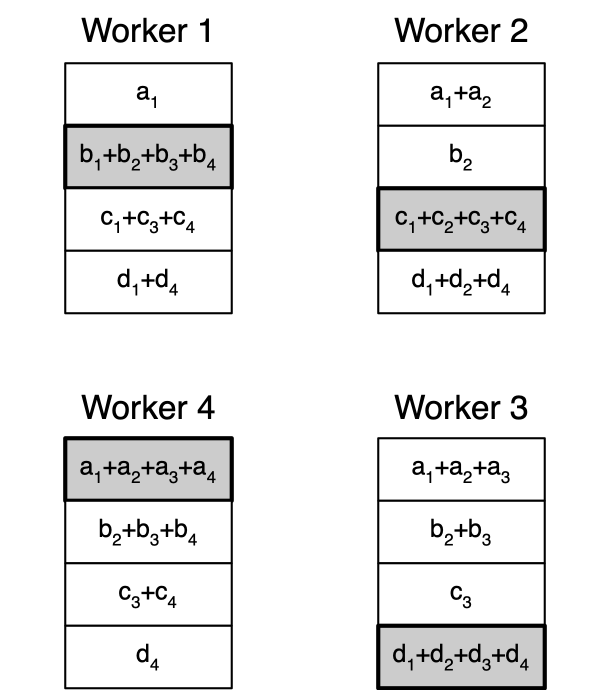}
\label{fig:Ringallreduce-4}}

\caption{Ring Allreduce: Reduce-Scatter phase.} 
\label{fig:ReduceScatter}
%

\centering 
\subfloat[]{\includegraphics[width=0.24\textwidth]{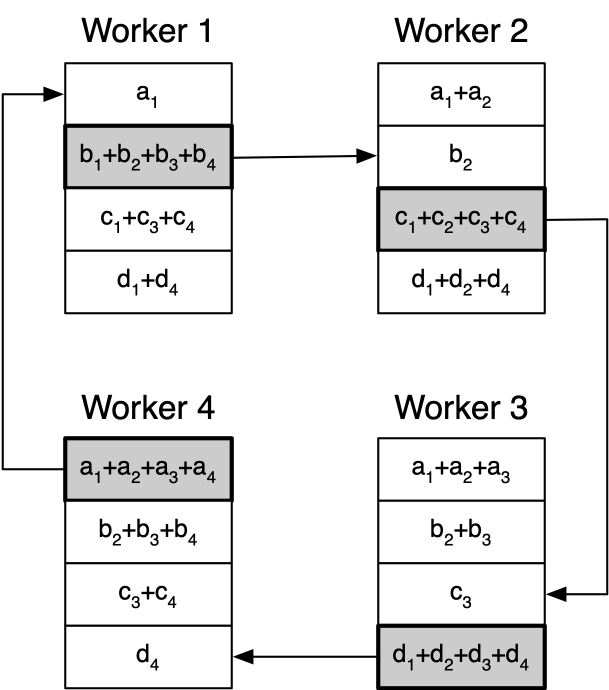}
\label{fig:Ringallreduce-5}}
\subfloat[]{\includegraphics[width=0.24\textwidth]{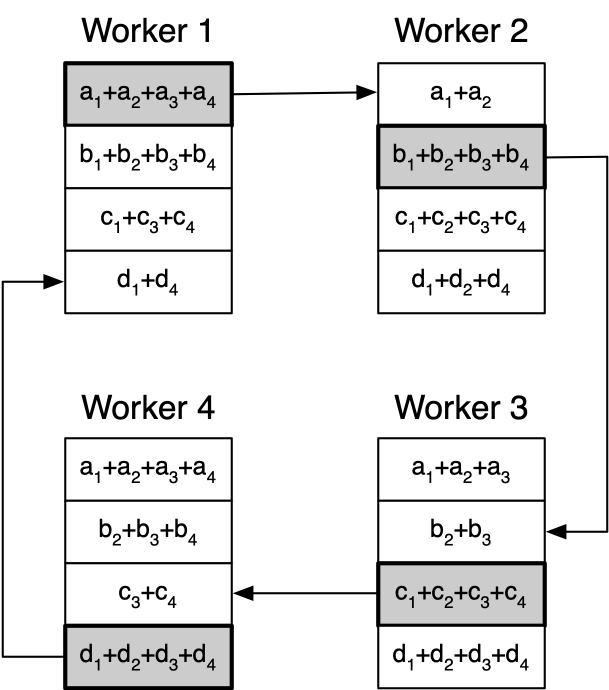}
\label{fig:Ringallreduce-6}}
%
%
\subfloat[]{\includegraphics[width=0.24\textwidth]{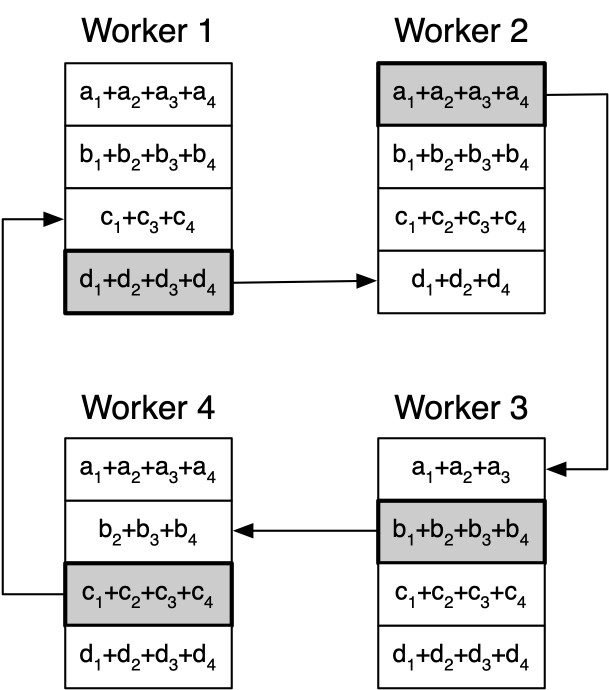}
\label{fig:Ringallreduce-7}}
\subfloat[]{\includegraphics[width=0.24\textwidth]{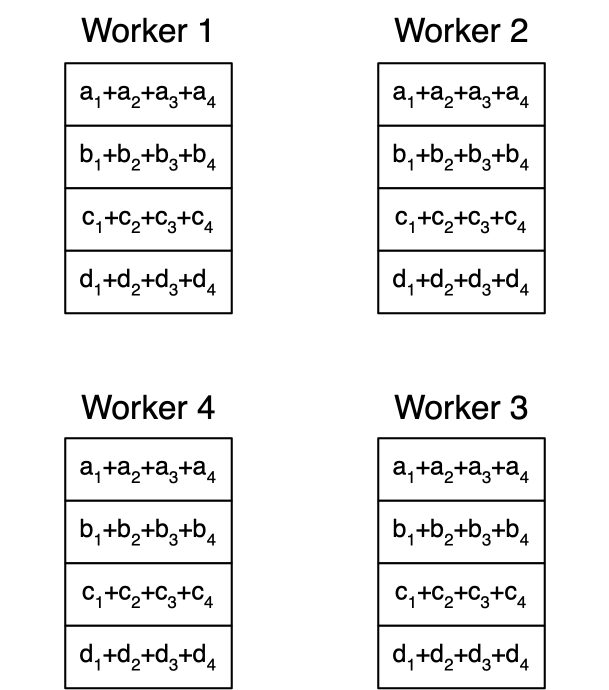}
\label{fig:Ringallreduce-8}}

\caption{Ring Allreduce: Allgather phase.} 
\label{fig:Allgather}
\end{figure}
%

\subsection{Ring AllReduce Algorithm}
\label{sec:rw_ring_explained}

In a DML setting, the AllReduce communication primitive 
computes the sum or average of gradient tensors, where each tensor is held 
by a different worker node. 
Generally, one AllReduce operation must 
be performed for most model layers in each round of training. 
In Ring Allreduce, workers are organized in a logical ring. Each tensor is divided into $N$ equal-sized chunks, where $N$ is the number of workers. 
The algorithm proceeds in two rounds of exchanges: {\it a Scatter-Reduce} phase, at the end of 
which each worker holds the global sum of one chunk, followed by an {\it All-Gather} 
phase that circulates those  chunks around the ring until
every worker holds the complete result. 
Both phases involve communication steps, where workers send data to their 
successor in the logical ring. 

\paragraph{Scatter-Reduce phase.} In each communication step,  each worker sends 
one chunk of the tensor data to its successor in the ring. A worker that receives a 
chunk adds it element-wise to its locally stored chunk and forwards the result  
to its own successor in the next communication step.
After $N-1$ communication steps, each worker holds the
reduced value, that is, the global sum, of one chunk.

Fig.~\ref{fig:ReduceScatter} illustrates the Scatter-Reduce phase for
$N=4$ workers, where each Worker$_j$ ($j=1,2,3,4$) initially holds a one-dimensional vector
$\{a_j, b_j, c_j, d_j\}$. Here, each element of the vector represents a chunk. 
The Reduce-Scatter phase starts by having each worker 
send a different element to its successor in the ring (Fig.~\ref{fig:Ringallreduce-1}). The receiver of an element adds it to its own  and then sends the result to its own successor (Fig.~\ref{fig:Ringallreduce-2}). After three steps, each worker holds the sum over all workers for one element  (Fig.~\ref{fig:Ringallreduce-4}). 

\paragraph{All-Gather phase.} In this phase, the reduced chunks are  disseminated across all workers. 
As before, this is done by having workers send data to their successor in the ring. 
Each worker transmits a reduced chunk to the next worker,  replacing the currently held chunk 
at the receiver.
After $N-1$ such steps, each worker holds the reduced value of 
each chunk. 
Figs.~\ref{fig:Ringallreduce-5}--~\ref{fig:Ringallreduce-7} illustrate the dissemination for four workers, and 
Fig.~\ref{fig:Ringallreduce-8} presents the final result. 

The steps  of Scatter-Reduce and All-Gather introduce local synchronization barriers whereby 
a worker can start the next step only if it has received data sent by its predecessor in the current step. Hence, if 
computing and communication times at workers are time-varying, workers may be forcibly idle, waiting for data to arrive 
from their predecessors. On the other hand, Ring AllReduce does not involve a global synchronization barrier. In particular, 
when a node has received the data chunk in the last step of Reduce-Scatter, it can immediately proceed with the first step of 
the All-Gather phase.

\subsection{Analytical Models of Ring AllReduce}
\label{sec:rw_modeling_hard}

The vast majority of performance analyses of the Ring AllReduce algorithm  
is focused on  communication overhead, generally in a deterministic setting. 
A frequently used  framework for determining collective communication delays is the
latency--bandwidth model introduced by Hockney~\cite{Hockney1994}, which
expresses the communication delay between two nodes in terms of a startup
latency~$\alpha$ and a per-byte transfer time~$\beta$. 
For Ring AllReduce, the model has been extended to account for reduction operations in the Reduce-Scatter phase~\cite{Thakur2005,Chan2007}. 
Under  a deterministic  model, the completion time 
for a Ring AllReduce operation for a gradient tensor of size $M$~bytes with $N$ workers takes 
the form  
\begin{align*}
2(N-1) \left(\alpha + \beta \tfrac{M}{N} \right) + (N-1) \gamma \tfrac{M}{N} \, , 
\end{align*}
where $\alpha$ is the startup latency for a transmission, $\beta$ is the transmission time 
per byte, and $\gamma$ is the computation time per byte for the reduction operation. 
Variations of this formula have been used in~\cite{Barrachina2023,Naumov2020,Nuriyev2021,Nuriyev2024SUARA,AlqahtaniDemirbas2019,alqahtani2019performance}.
Alqahtani and Demirbas~\cite{AlqahtaniDemirbas2019} present analytical
performance models for parameter-server, peer-to-peer, and Ring AllReduce architectures. 
They derive deterministic expressions for computation and communication time as a function of
model size, batch size, and the number of workers, and validate 
results using measurements of TensorFlow~\cite{tensorflow2015-whitepaper} and Horovod~\cite{ringallreduce-uber}. Since deterministic models do not account for 
delay variations in computation and communication, they are not suitable for 
capturing idle times of workers due to synchronization barriers, resource contention, or 
heterogeneity of computing resources. 

Consideration of network resource contention and runtime variability 
among workers, that cause synchronization delays,  calls for a  
stochastic analysis. 
However, the dependency structure imposed by the logical ring 
makes such an analysis challenging.  
An analysis of gradient reductions in~\cite{li2024straggler} models the 
completion times of Ring Allreduce at workers by independent random variables and characterizes the completion of the collective through
their maximum. This formulation captures variability in worker completion
times, but does not capture the propagation of delays through the stepwise
dependencies of the ring, where workers wait for data required to continue communication.

Chen et al.~\cite{chen2024rina} identify the dependencies between workers as 
a source of performance degradation in Ring AllReduce, where  a single 
slow worker impacts completion times for  the entire
collective. Assuming Gaussian delays for computation and communication, they present an expression 
for the average completion time of Ring AllReduce. 
Their analysis  assumes a global synchronization barrier after each step 
of Scatter-Reduce and All-Gather. However, Ring AllReduce  relies exclusively on local 
 synchronization between neighbors in the logical ring, without global synchronization. 
The assumption of global barriers simplifies the analysis as it decouples existing dependencies between workers across multiple 
steps. At the same time, global synchronizations overestimate the times that workers are idle, resulting in overly pessimistic 
estimates of completion times. 

In~\cite[Appendix C]{SDRRDMA2028}, the authors derive a lower bound on the expected time of a Ring AllReduce operation 
using random delays in each step of the algorithm. Assuming that the delay of each step is 
given by the sum of a constant $C$ and a random variable $X$, the expected value of the lower bound 
for $N$ workers is given as $2(N-2) (C+E[x])$. The analysis 
isolates the multiple steps of Ring AllReduce 
via an application of Jensen's inequality.

Yu et al.~\cite{yu2022scheduling,Yu2022GADGET} study the impact of contention at a communication bottleneck 
on completion times of Ring AllReduce in a GPU cluster. They formulate resource scheduling at a bottleneck 
resource 
as a non-linear integer program, which is solved using suitable heuristics. 
Their results show that the communication
delays in Ring AllReduce job depends not only on gradient size and  link
bandwidth, but also on contention for shared communication resources.

Using measurement experiments, Zhang and
Shen~\cite{zhang2026straggler} show that computation and 
communication times in distributed training can exhibit large variations even in
a homogeneous GPU cluster. Measurement experiments in~\cite{li2024straggler} also find  
substantial variations in the
network bandwidth available to workers under resource contention in a homogeneous 
environment. The authors conclude that a single deterministic service rate cannot
represent this variation, and that stochastic models are needed to characterize the
distribution of worker service times and allow these distributions to differ
across workers. Assuming independent and identically distributed gradient 
transmissions, they derive the expected completion time of 
Ring AllReduce.

So far, there does not exist a stochastic analysis that tracks the dependencies between $N$~workers over all $2(N-1)$~steps of Ring AllReduce. 
The analysis presented in this paper seeks to fill  this gap. By presenting models that represent dependencies between 
workers across all steps of Ring AllReduce, we obtain accurate 
estimates of the fraction of idle workers 
and the completion time of Ring AllReduce in a stochastic regime.

%% file: 3_Modeling_Serverless_Distributed_Training_Using_SPN_Nov23.tex
\section{Modeling Serverless Distributed Training Using SPN}
\label{sec:spn-model}

In this section,  we construct a probabilistic model of   Ring AllReduce 
that allows us to study the impact of random variations in serverless collective communications. 
In our model, the communication and computation times at a worker in each step of
Ring AllReduce are viewed as governed by random variables. 
Our approach seeks to complement  empirical studies of communication and computation overhead in
serverless ML systems~\cite{Copik2023FMI,barrak2023spirt} 
and the variability induced by architectural features of serverless
platforms~\cite{GimenoSarroca2024MLLess,besozzi2026hpcserverless}. 
As the existing empirical studies, we seek to provide explanations for the observed idle times 
in serverless collective communications. Different from previous works, we do not 
investigate the impact of subtasks, such as memory access, data processing,  or process 
scheduling. Instead, we explore randomness of computation and communication as root cause of low resource utilization.
In this section, we capture  concurrency and synchronization barriers of the Ring AllReduce operation in the presence of randomness 
using a stochastic Petri net (SPN) model.

\begin{figure}[!t]
\centering 
\subfloat[Before firing.]{\includegraphics[width=1in]{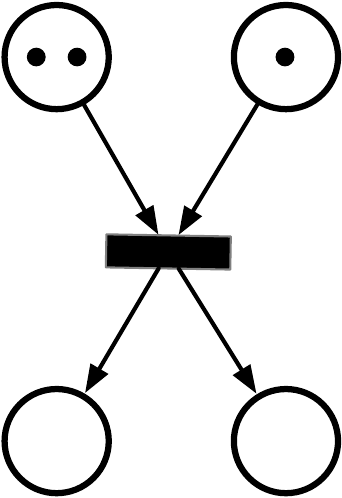}\label{fig:pn1}}
\hspace{3cm}
\subfloat[After firing.]{\includegraphics[width=1in]{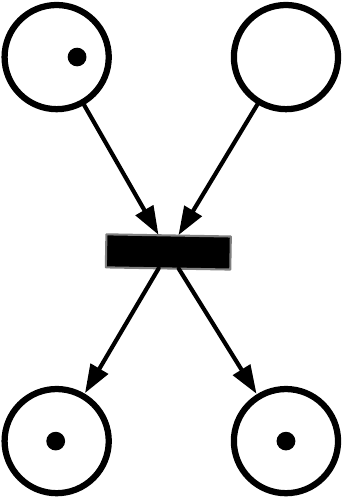}\label{fig:pn2}}
\vspace{-5pt}
\caption{Firing of a transition in a Petri net. A transition can fire when each input place contains at least one token. The firing of a transition removes one token from each input place and adds one token to each output place.} \label{fig:PN}
\end{figure}

\subsection{Short Review of Petri Nets}
\label{sec:spn_background}

A Petri Net (PN)~\cite{peterson1981petri} is a directed graph with two types of nodes, called {\it places} and {\it transitions}, used to model a distributed discrete-event system with  concurrency and synchronization.
All edges of the graph are  between places and transitions. 
A place with an edge to a transition is called an {\it input place} and a place with an 
edge from a transition is called an {\it output place}. 
Places are labeled with tokens, and the number of tokens in each place, referred to as {\it marking}, indicates the state of the PN. The dynamics of a PN are represented by moving tokens between places. 
If each input place of a transition contains one or more tokens, 
the transition is said to be {\it enabled} and  allowed to {\it fire}. During {\it firing}, one token is removed from each input place, and one token is added to each output place.
For illustration, Fig.~\ref{fig:pn1} depicts a PN with four places, indicated by circles, and one transition, indicated by a rectangle.  Tokens are represented by black dots. 
Since the input places of the transition both contain tokens, the transition is enabled and can fire.  
Fig.~\ref{fig:pn2} shows the marking after the firing of the transition. 
Note that in Fig.~\ref{fig:pn2}, the transition is no longer enabled. 
An SPN is an extension that associates each
transition with an exponentially distributed firing delay. With exponential firing delays, the set of feasible markings is equivalent to a 
CTMC~\cite{Marsan1995SPN}.

\subsection{SPN model of Ring AllReduce}
\label{sec:spn_model}

We next develop an SPN representation of Ring AllReduce for $N$ workers. 
The SPN model encompasses the steps of the Reduce-Scatter and All-Gather phases discussed in Sec.~\ref{sec:rw_ring_explained}. Recall that in a Ring AllReduce with $N$ workers, there are $2(N-1)$ steps that each involve a local computation and a transmission of a data chunk to the next worker in the ring. A worker that has sent a data chunk to its successor  in the ring can proceed only after receiving the next chunk  from its predecessor.  

\begin{figure}[t!] 
\centering
\includegraphics[width=3in]{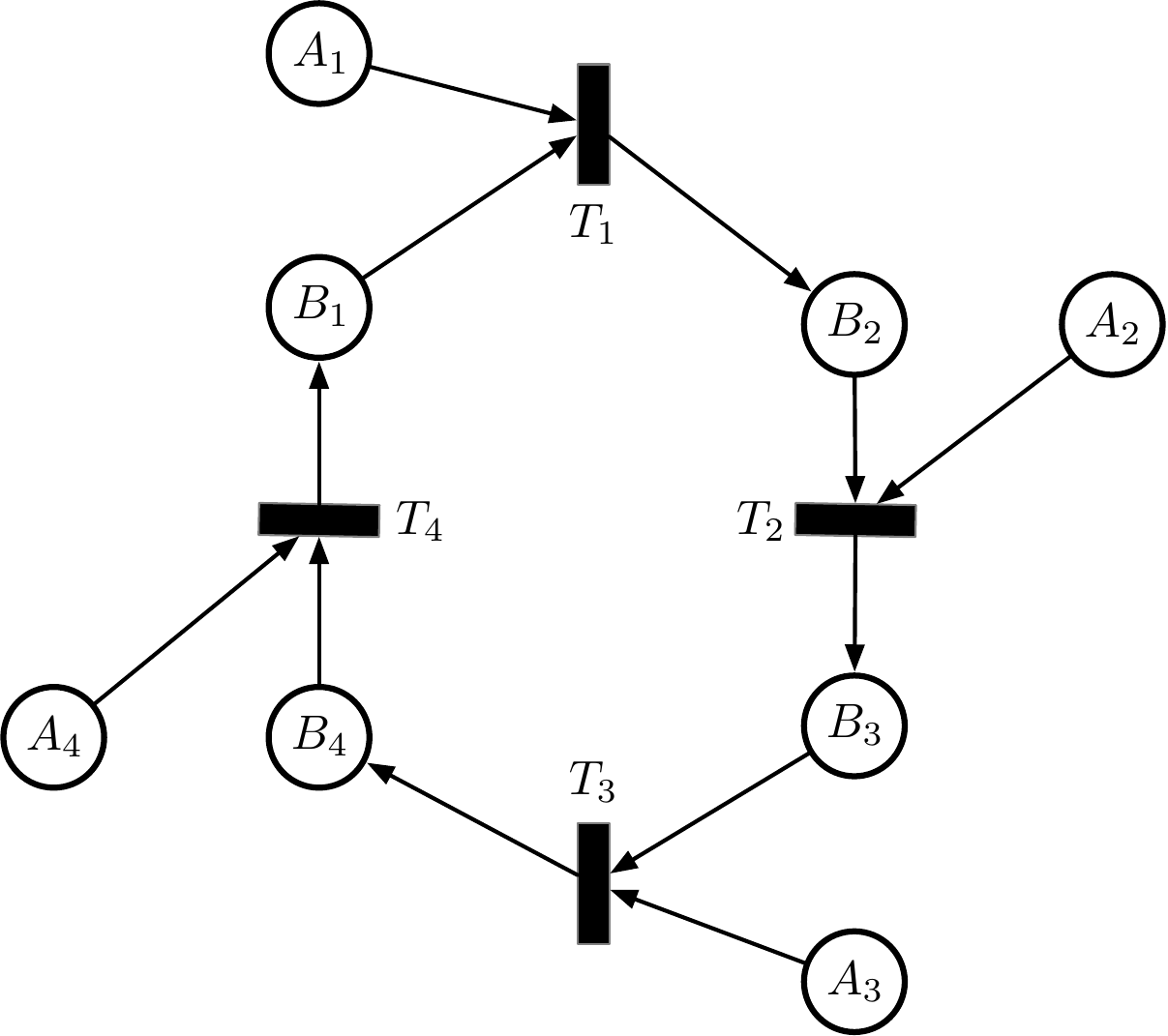}
\caption{SPN model of Ring AllReduce with four workers.}
\label{fig:spn}
\end{figure}

Fig.~\ref{fig:spn} illustrates an SPN model of Ring AllReduce for $N=4$ workers in a clockwise arrangement.  
For ease of notation,  indices are interpreted modulo~$N$. With this, $\text{Worker}_{N+1} \equiv \text{Worker}_1$  
and $\text{Worker}_{0} \equiv \text{Worker}_N$. 

For each Worker$_i$, there are two places $A_i$ and $B_i$, and one transition $T_i$. 
Transition $T_i$ has  $A_i$ and $B_i$ as input places
and  $B_{i+1}$ as  output place. Transition $T_i$ is enabled
when both $A_i$ and $B_i$ contain at least one token.
We will use $n_{A_i} (t)$ and $n_{B_i} (t)$, respectively, to denote 
the marking of $A_i$ and $B_i$ at time $t$, with $t=0$ indicating the start of 
the operation. 

\begin{itemize}
  \item Place $A_i$ keeps track of the remaining steps to be performed by Worker$_i$. The initial marking is $n_{A_i} (0)= 2(N{-}1)$, that is, there is one token for each step of  Reduce-Scatter and All-Gather.  

  \item Place $B_i$, with initial marking $n_{B_i} (0)= 1$ represents the communication state of Worker$_i$ with respect to its predecessor in the ring. If $n_{B_i} (t)= 0$,  Worker$_i$ is blocked waiting for a data chunk  
 from its predecessor. 
 
  \item Transition $T_i$ represents the delay incurred in one step of the Ring AllReduce operation at  Worker$_i$. It includes processing a data chunk and forwarding the
  result to Worker$_{i +1}$. 
  $T_i$ is enabled when both $A_i$ and $B_i$ contain at least one token, that is, when Worker$_i$ has at least one pending step ($n_{A_i} (t)> 0$) and is not blocked waiting for a data chunk from the ring ($n_{B_i} (t)> 0$). 
  In the SPN model, each transition $T_i$ has an exponentially distributed firing delay with rate $\mu_i$. 
  Later, we will relax the assumption and consider firing delays that have a constant and a (exponentially distributed) random component. 
\end{itemize}

The execution of Ring AllReduce is captured
by the token dynamics. Initially, each worker has $2(N{-}1)$ remaining
steps represented in $A_i$ and one data chunk available in $B_i$. As
transitions fire, tokens in $A_i$ are consumed, while tokens in $B_i$
move around the ring, following the chunk-passing pattern of the
Reduce-Scatter and All-Gather phases shown in
Figs.~\ref{fig:ReduceScatter} and~\ref{fig:Allgather}. 
The absorbing marking in which all places $A_i$ are empty represents
the completion of the collective operation.
This occurs at the earliest time $t$ when $n_{A_i} (t)= 0$ and $n_{B_i} (t)= 1$ for all workers.

\subsection{Structural Invariants}

We next explore the structure of the SPN model. 
Using $Z_i(t)$ to denote the number of firings of transition~$T_i$ up to time $t$, the token count in 
places $A_i$ and $B_i$ at time $t$ evolves as 
\vspace{-1.5mm}
\begin{align}
    n_{A_i}(t) &= 2(N-1) - Z_i(t) \, ,  \label{eq111_simplified} \\
    n_{B_i}(t) &= 1 + Z_{i-1}(t) - Z_i(t) \label{eq222_simplified} \, . 
\end{align}

Our first observation, stated in Lemma~\ref{lemma1}, is that the token occupancy at place $B_i$ derives from the 
progress at Worker${_i}$ relative to that of Worker$_{i-1}$. 
\begin{lemma}\label{lemma1}
In the SPN model for Ring AllReduce, the number of tokens in each place \( B_i \) at time \( t \) is given by 
\begin{align}
n_{B_i}(t) = 1 + n_{A_i}(t) - n_{A_{i-1}}(t) \, . 
\label{eq:lem1-1}
\end{align}
\end{lemma}
\begin{proof}[\textbf{Proof}]
Based on Eq.~\eqref{eq111_simplified}, $Z_i(t)$ and $Z_{i-1}(t)$ are computed as
\begin{align}
    Z_i(t) &= 2(N-1) - n_{A_i}(t), \label{eq111_simplified_Z_i}\\
    Z_{i-1}(t) &= 2(N-1) - n_{A_{i-1}}(t). \label{eq222_simplified_Z_i_1}
\end{align}
Then, rewriting Eq.~\eqref{eq222_simplified} yields the result 
\begin{align*}
    n_{B_i}(t) &= 1 + \left(2\left(N-1\right) - n_{A_{i-1}}\left(t\right)\right)
     -\left(2\left(N-1\right) - n_{A_i}\left(t\right)\right) \\
    &= 1 + n_{A_i}\left(t\right) - n_{A_{i-1}}\left(t\right) \, . 
\end{align*}
 \end{proof}
\noindent 
The next lemma asserts that the total number of tokens across all $B_i$ places remains 
fixed. 
\begin{lemma}\label{Sum_N_b}
In the SPN model for Ring AllReduce, the total number of tokens across
all $B_i$ places is always equal to the number of workers. That is, for all times $t \ge 0$,  
\begin{align}\label{eq:lem1-2}
    \sum_{i=1}^{N} n_{B_i}(t) = N \, .
\end{align}
\end{lemma}

\begin{proof}
With Lemma~\ref{lemma1} and summing over all workers  yields
\begin{align}
    \sum\limits_{i=1}^{N} n_{B_i}(t) &= \sum\limits_{i=1}^{N} \left( 1 +n_{A_i}\left(t\right) - n_{A_{i-1}}\left(t\right) \right) \nonumber \\
    &= N + \sum\limits_{i=1}^{N} \left(  n_{A_i}\left(t\right) - n_{A_{i-1}}\left(t\right) \right) \nonumber\\
    &\stackrel{(a)}{=} N+ n_{A_N}(t) - n_{A_0}(t) \nonumber\\
    &\stackrel{(b)}{=} N \, ,\nonumber
\end{align}
where $(a)$ follows from the telescoping property of the sum, under
which the intermediate terms cancel, and  $(b)$ follows from the
cyclic indexing of the ring, where $A_0 \equiv A_N$.  
\end{proof}

The following lemma states that  
 the number of steps by which one worker can advance or fall behind relative 
to any other worker is bounded by the total number of workers. 
\begin{lemma}
\label{lem:nA_bound}
For all $i, j \in \{1,\ldots,N\}$ and $t \geq 0$, the token counts in
places $A_i$ and $A_j$ of the SPN model for Ring AllReduce satisfy
\begin{equation}\label{eq:nA_diff_bound}
    |n_{A_i}(t) - n_{A_j}(t)| \;\leq\; N - 1 \, .
\end{equation}
\end{lemma}

\begin{proof}[\textbf{Proof}]
If $i \equiv j \pmod{N}$, the left-hand side of
Eq.~\eqref{eq:nA_diff_bound} is zero and the inequality holds trivially.
Otherwise, by the symmetry of the absolute value it suffices to show
that $n_{A_i}(t) - n_{A_j}(t) \leq N - 1$ for an arbitrary ordered
pair $(i, j)$.

Recalling that indices are taken
modulo $N$, let $i$ and $j$ be two distinct indices. The clockwise path from
place~$A_j$ to $A_i$ is the  sequence of traversed transitions 
when moving clockwise along the ring from $A_j$ to $A_i$. The number
of steps on this path is
\begin{align*}
    k := (i-j)\bmod N \quad \text{with} \ k \in \{1,\ldots,N-1\} \, .
\end{align*}
The path visits the sequence
$j=\ell_0 ,\,\ell_1,\ldots,\ell_k = i$, where
\[
    \ell_m = j+m\mod N \, ,
    \qquad m=0,1,\ldots,k \, .
\]
Applying the telescoping identity along this path we derive 
\begin{equation}
    \label{eq:path_sum}
\begin{split}
    n_{A_i}(t) - n_{A_j}(t)
    &= \sum_{m=1}^{k}
       \bigl(n_{A_{\ell_m}}(t) - n_{A_{\ell_{m-1}}}(t)\bigr) \\
    &\stackrel{(c)}{=} \sum_{m=1}^{k} \bigl(n_{B_{\ell_m}}(t) - 1\bigr) \\
    &= \sum_{m=1}^{k} n_{B_{\ell_m}}(t) \;-\; k \, ,
\end{split}
\end{equation}
where $(c)$ applies the identity
$n_{B_{\ell_m}}(t) = n_{A_{\ell_m}}(t) - n_{A_{\ell_{m-1}}}(t) + 1$
from Lemma \ref{lemma1}. Since $i \neq j$ implies $k \geq 1$, it follows from
\eqref{eq:path_sum} that
\begin{align*}
    n_{A_i}(t) - n_{A_j}(t)
    \;\stackrel{(d)}{\leq}\; \sum_{\ell=1}^{N} n_{B_\ell}(t) - 1 
    \;\stackrel{(e)}{=}\; N - 1 \, ,
\end{align*}
where $(d)$ follows from extending the partial sum over all $N$
workers, and $(e)$ applies Lemma~\ref{Sum_N_b}.
\end{proof}

We next provide a renewal relation that 
governs the sequence of firings of the transitions  $T_i$.
For $0\le r\le 2(N-1)$, let $\tau_{i,r}$ denote the time
when Worker$_i$ completes the $r$-th firing.
By convention, we set $\tau_{i,0}=0$. 
For $r>0$, $\tau_{i,r}$ 
can be expressed in terms of the variables introduced in the proof of 
Lemma~\ref{lemma1} as
the stopping time
\begin{equation}
\label{eq:tau-ir}
\tau_{i,r} = \sup \bigl\{t\ge 0\ \big\vert\ Z_{i}(t)<r\bigr\}.
\end{equation}
We emphasize that the renewal relation
depends on the initial markings 
but requires no assumptions on the firing delays.

\begin{lemma}\label{lemma:stopping_time}
In the SPN model for Ring AllReduce, 
denote by $S_{i,r}$ the $r$-th firing delay of Worker$_i$,
and let $\tau_{i,r}$ be the stopping times
of Worker$_i$ defined above. Then
\begin{align}
\label{eq:renewal}
\tau_{i,r+1} = 
\max \left\{ \tau_{i,r}, \tau_{i-1,r}\right\}
+ S_{i,r+1}
\end{align}
holds for all $r\ge 0$ such that 
$0\le r <\min_{i} n_{A_i}(0)$.
\end{lemma}

\begin{proof}
Transition $T_i$ is enabled for the ($r+1$)-st 
time as soon as $Z_i(t)=r$ and $n_{B_i}(t)>0$,
since $n_{A_i(t)}= n_{A_i}(0)-r>0$ by assumption.
By Eq.~\eqref{eq222_simplified}, this
requires $Z_{i-1}(t)\ge n_{B_i}(t) + Z_i(t)-1 = r$.
By definition of the stopping times, 
this first happens at time $\max \left\{\tau_{i,r}, \tau_{i-1,r}\right\}$.
Since $r<2(N-1)$, at that time we also have $n_{A_i}(t)>0$ 
by Eq.~\eqref{eq111_simplified}.
	Once the transition is enabled, the $(r+1)$st firing is completed 
after the delay $S_{i,r}$.
\end{proof}

\subsection{Markov Chain Analysis}
\label{subsec:CTMC}

We next explore the state space of the Markov chain that is induced by the SPN 
model.  Each reachable marking represents a distinct state. We represent a state by the token-count vector
\[
\mathbf{n} = \left(n_{A_1}, n_{B_1}, \dots, n_{A_N}, n_{B_N}\right)\, , 
\]
where $n_{A_i}$ and $n_{B_i}$, respectively, are the markings of 
places $A_i$ and $B_i$. 
The Markov chain
is started at time $t=0$, and its
initial state is given by
$n_{A_i} (0)= 2(N-1)$ and $n_{B_i} (0)=1$ for all $i\in \{1,\ldots, N\}$.
We let $\mathcal{S}$ denote the set 
of reachable states. 
The state of the Markov chain at time $t$ is denoted by  $\mathbf{n}(t)$.

Each firing of a transition in the SPN corresponds to a state 
transition in the Markov chain.  
The set of enabled transitions in a state $\mathbf{n} \in \mathcal{S}$ is given by 
the index set  
\[
    \mathcal{E} (\mathbf{n}) \;=\;
    \bigl\{i \in \{1,\ldots,N\} \; \mid \;
    n_{A_i} \geq 1 \text{ and } n_{B_i} \geq 1\bigr\},
\]
Each enabled transition $T_i$, $i \in \mathcal{E}(\mathbf{n})$, incurs an 
exponentially distributed firing delay with rate $\mu_i$. 
Note that with probability 1, only one transition fires at a time.  

Upon firing of $T_{i}$ at some time~$t$, the state of the Markov chain is updated as
\begin{equation}
\begin{split}
    n_{A_{i}}   (t)  &\;\longleftarrow\; n_{A_{i}} (t^-) - 1, \\
    n_{B_{i}}   (t)  &\;\longleftarrow\; n_{B_{i}} (t^-)- 1, \\
    n_{B_{i+1}} (t)  &\;\longleftarrow\; n_{B_{i+1}} (t^-)+ 1,
\end{split}
	\label{eq:MC-update}
\end{equation}
with $n_{A_{j}} (t)$ and $n_{B_{j}} (t)$ unchanged for all $j \neq i$. Here we use 
$t^-$ to denote the time instants just before time~$t$. 
The Markov chain has an absorbing state, which corresponds to the completion of the AllReduce operation, where $n_{A_i} = 0$ and $n_{B_i}=1$ for all~$i$. 

We are mainly interested in  two performance metrics of the Markov chain. 
\begin{enumerate}
\item {\bf Forced idle probability.} A forced idle time refers to a time period when a worker has unfinished work but is waiting for a data chunk from its predecessor in 
the ring. For Worker$_i$, this refers to all states where 
 $n_{A_{i}}> 0$ and $n_{B_{i}} =0$. We measure the forced idle 
 times in terms of the probability that a worker is forcibly idle. 
We refer to this probability as forced idle (or blocking) probability and 
denote it by ${\mathsf P}_{{\rm idle},i}$ for Worker$_i$. 
 If all workers are homogeneous, the probabilities are identical and we drop the 
 subscript. 
 
 \item {\bf Completion Time.} The completion time
refers to the total duration of the 
Ring AllReduce operation from start to finish.
In the Markov chain, this corresponds to the elapsed time between the initial state $(2(N-1), 1, \dots, 2(N-1), 1 )$ to the absorbing state $(0, 1, \dots, 0, 1 )$. 
		Let us denote the completion time of Worker$_i$ by $T_{{\rm RAR}, i}$\footnote{`RAR' stands for {\bf R}ing {\bf A}ll{\bf R}educe.} with 
\begin{align}   
\label{eq:T-rar-i}
T_{{\rm RAR},i} = \inf \{t \, \mid \, n_{A_i} (t) = 0 \} \, ,
\end{align}
and by $T_{\rm RAR}$ the completion time of Ring AllReduce. 
Since Ring AllReduce is completed when the last worker 
finishes we have 
\begin{align}   \label{eq:T-rar}
T_{\mathrm{RAR}} \;=\;  \max_{i=1,\ldots, N} 
T_{{\rm RAR},i} \, . 
\end{align} 
Note that in the SPN, the completion times $T_{{\rm RAR},i}$ 
and $T_{{\rm RAR}}$ 
are random variables. 
\end{enumerate}

In principle, the above performance metrics  can be obtained 
by a standard Markov-chain analysis.  
However, this approach does 
not scale, since the size of the reachable state space 
grows exponentially as $4^{N}$ for $N$ workers. 
As a result, 
a numerical solution becomes infeasible even for a modest number of workers. 
Moreover, by Lemma~\ref{lemma1}
the token-count vector 
$\mathbf{n} = \left(n_{A_1}, n_{B_1}, \dots, n_{A_N}, n_{B_N}\right)$
contains redundant information, since 
the markings $n_{B_i}$ are determined by
the markings $n_{A_i}$ through Eq.~\eqref{eq:lem1-1}.
Conversely, $n_{A_2},\ldots n_{A_N}$
can be computed from
$n_{B_1},\ldots, n_{B_N}$ and $n_{A_1}$ by
solving the recursion in Eq.~\eqref{eq:lem1-1}.
This motivates the investigation of 
approaches that avoid constructing the full Markov chain.

\begin{figure}[t!]
    \centering
    \includegraphics[width=1\linewidth]{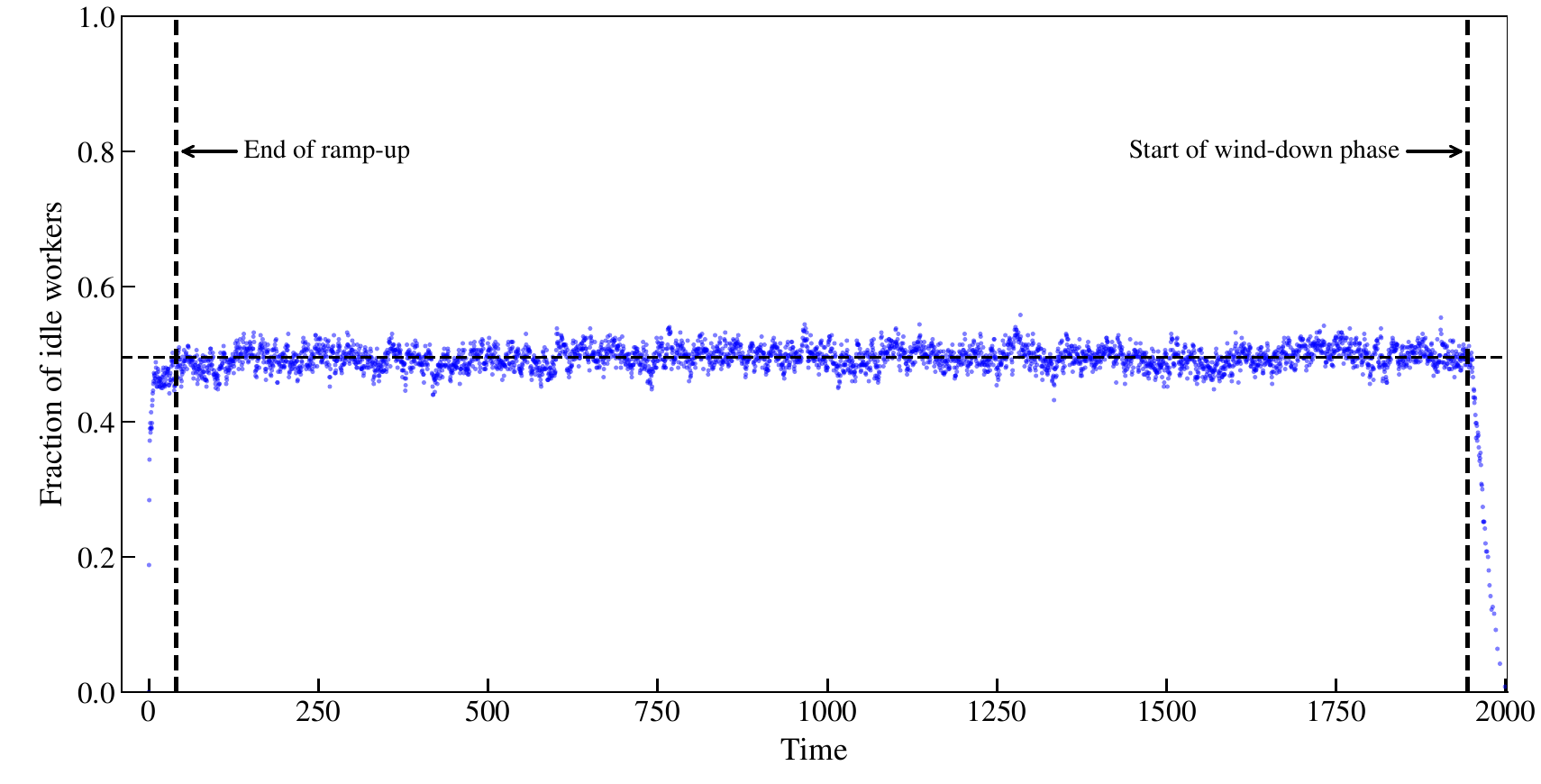}
    \caption{Fraction of idle workers in Ring AllReduce SPN model with $N = 500$
    workers. Firing delays are exponentially distributed with rate $\mu=1$ for all workers. The dashed vertical lines
    separate the three phases: ramp-up, stable, and wind-down.}
    \label{full_run_transient}
\end{figure}

Before presenting our analysis, we first discuss a simulation of the CTMC for Ring AllReduce 
with $N=500$ workers and rate $\mu = 1$ for each worker. 
Fig.~\ref{full_run_transient} shows the fraction of forcibly idle workers over the execution of Ring AllReduce.  A finished worker, that is,  a Worker$_i$ with $n_{A_i}=0$, 
is not considered idle.  
The  figure plots the fraction of idle workers for every (round) time unit, i.e., at times $1,2,3, \ldots$.

The graph shows that the Ring AllReduce operation has three distinct phases. 
Most of the time, the idle fraction hovers around a value close to $\tfrac{1}{2}$, meaning 
that almost half the workers are idle. We refer to this phase as {\it stable state}.\footnote{Since the Markov chain is absorbing we avoid the term `steady state'.} 
The stable state is preceded by a {\it ramp-up} where the fraction of 
idle workers increases and followed by a {\it wind-down} where workers 
are completing their computations.  
The boundaries between the 
phases are indicated by dashed lines.
The boundary between the stable state and the wind-down, 
at time $T_{\mathrm{wind-down}}$, is defined as the time when the first 
worker finishes its Ring AllReduce work is, that is, 
\begin{align}   \label{eq:T-wd}
	T_{\mathrm {wind-down}} \;=\;  \min_{i=1,\ldots, N}  
T_{{\rm RAR},i} \, . 
\end{align} 
The delineation between  ramp-up and stable state is not 
as clearly defined. In Fig.~\ref{full_run_transient} we use  
the Marginal Standard Error 
Rule (MSER)~\cite{White1997MSER,WhiteCobbSpratt2000} 
applied to the samples of the fraction of idle workers collected before~$T_{\mathrm{wind-down}}$. 

The simulation highlights several aspects of the Ring AllReduce operation with a large number of workers. 
Foremost, with a relatively short ramp-up and wind-down, the operation is largely characterized 
by a stable state with a relatively constant 
idle probability. This justifies a steady-state analysis of the SPN to obtain a first-order estimate of performance metrics. In the next section, 
we derive expressions for the forced idle probabilities and completion times  
of Ring AllReduce under different scenarios.


%% file: 4_Analysis_of_Idle_Times_Nov23.tex
\section{Analysis of Forced Idle Probabilities and Completion Times}
\label{sec:idle_time_analysis}
This section develops an analytical characterization of worker idle times
in Ring AllReduce. As seen in the example in Subsec.~\ref{subsec:CTMC}, workers are idle
for a non-negligible fraction of the execution time, even when all
workers have identical service rates. Since a direct solution 
of the Markov chain does not scale, we
exploit the structural properties of the SPN to obtain analytical results. 
In Subsec.~\ref{subsec:Homo-Exp}, we consider that all workers 
incur  delays drawn from the same exponential distribution and 
obtain the forced idle probability with a simple stars-and-bars argument. 
We extend this argument to a model with limited 
randomness in Subsec.~\ref{subsec:Homo-Shift-Exp}. 
In Subsec.~\ref{subsec:pfqn}, we analyze a setting with heterogeneous workers 
through a queueing network model. This allows us to consider the impact of 
stragglers on the forced idle probability. 

Our objective in this section is to obtain the forced idle 
probability in the stable state, that is,  
away from the ramp-up and wind-down phases. 
For this, we consider tokens 
perpetually rotating through the ring topology, 
which is achieved by setting 
the initial markings of places $A_i$ to 
$n_{A_i} (0) \to \infty$ for each Worker$_i$. 
Since in this case, places $A_i$ have no impact on the 
firing of transitions, we will use a reduced SPN model 
that only has places $B_1, \ldots , B_N$ and transitions $T_1, \ldots ,T_N$. 
The state space of the reduced SPN is 
\begin{align}
\mathbf{n} = \left(n_{B_1},\ldots,n_{B_N}\right)
\label{eq:state-spn}
\end{align}
with $n_{B_i} \geq 0$ and $\sum_{i=1}^{N} n_{B_i} = N$. 
A firing of transition~$T_{i}$ triggers a state change 
where $n_{B_{i}} > 0$ is decremented and $n_{B_{{i+1}}}$ is incremented. 

\subsection{Exponential Firing Delays}
\label{subsec:Homo-Exp}

We consider the reduced SPN model of Ring AllReduce with $N$~workers described above, with initial marking 
$n_{A_i} (0)\to\infty$ and  $n_{B_i} (0)=1$ for $i = 1, \ldots, N$, 
where firing delays at all workers are exponentially distributed with rate $\mu$. 
Under these assumptions, the SPN is a CTMC whose
stationary distribution is uniform over the reachable states. 

Denote by $\mathsf{P}_{\rm idle}(t)$ the forced idle probability
of any given worker at time $t$. Note that this
probability is the same for all workers.
Our first result shows that the
forced idle probability increases almost instantaneously
to a value above $0.3$, regardless of $N$.

\begin{lemma} For every $N\ge 2$,
the forced idle probability of each worker satisfies
\[
{\mathsf P}_{{\rm idle}}(t)\ge 0.3
\qquad \text{for all} \ t\ge \mu^{-1}\,.
\]
\end{lemma}

\begin{proof} 
For $j=0,\ldots, N$, denote
by $m_j(t)$ the number of workers at
time $t$ that hold exactly $j$ tokens.
By definition, 
$${\mathsf P}_{\rm idle}(t) = \frac1N \mathbb{E}[m_0(t)]\,.
$$

We will derive
and solve a differential
inequality for ${\mathsf P}_{\rm idle}(t)$.
By the property of the exponential firing
rates, if $n_{B_i}(t)=1$ then Worker$_i$
becomes newly blocked at rate
$\mu$, and if $n_{B_i}(t)=0$ it may 
become unblocked at rate $\le \mu$ (depending on
$n_{B_{i-1}}(t)$). Averaging over the workers, it follows that
\[
\frac{d}{dt} {\mathsf P}_{\rm idle}(t)
=\frac1N \frac{d}{dt} \mathbb{E}\left[m_0(t)\right]\\
\ge \frac{\mu}{N} \mathbb{E}\left[m_1(t)-m_0(t)\right]\,.
\]
To estimate the right hand side, we use that
\[
\sum_{j=0}^N m_j(t)=N\quad\text{and} \quad
\sum_{j=0}^N j m_j(t)=N,
\]
where the first equality expresses that
the number of workers equals $N$, and second
that the total number of tokens equals $N$
by Lemma~\ref{Sum_N_b}.
Subtracting the sums, we obtain
\begin{align*}
0 &= \sum_{j=1}^N (j\!-\!1) m_j(t)\\
& \ge -m_0(t) + \sum_{j=2}^N m_j(t)\\
&= N-2m_0(t)-m_1(t)\,.
\end{align*}
In the second line, we have
isolated the terms with $j=0, 1$, and used that
$j-1\ge 1$ for $j\ge 2$.
In the last line, we have recalled
that the number of workers is $N$.
It follows that $m_1(t)-m_0(t)\ge N-3m_0(t)$,
and hence
\[
\frac{d}{dt} {\mathsf P}_{\rm idle}(t)
\ \ge \ \frac{\mu}{N}\mathbb{E}\left[\bigl(N-3 m_0(t)\bigr)\right]
\ =\ \mu\bigl(1-3{\mathsf P}_{\rm idle}(t)\bigr)
\]
for all $t\ge 0$. 
We conclude from the differential
inequality and initial value ${\mathsf P}_{\rm idle}(0)=0$
that $\mathsf{P}_{\rm idle}(t)$ increases with $t$, and
\[
{\mathsf P}_{\rm idle}(t) \ge \frac13\left(1-e^{-3\mu t}\right)\,,
\qquad t\ge 0\,.
\]
The claim follows since 
${\mathsf P}_{\rm idle}(t) \ge 
{\mathsf P}_{\rm idle}(\mu^{-1}) 
= \frac13\left(1-e^{-3}\right) > 0.3$
for $t\ge \mu^{-1}$.
\end{proof}

We next give an exact expression for the idle probability in the stable 
state, which we obtain from a steady-state analysis. 
Throughout the paper, we indicate steady-state performance 
metrics with a star ($\ast$). 
The steady-state forced idle probability ${\mathsf P}^\ast_{\rm idle}$, 
determined in the following proposition, follows from a stars-and-bars
count~\cite{flajolet2009analytic}. 

\begin{proposition}[Steady-state Idle Probability under Pure-exponential Firing Delays] 
\label{lem:pblock}
Consider the SPN model of Ring AllReduce from Fig.~\ref{fig:spn} with $N$~workers, with initial marking 
$n_{A_i} (0)=\infty$ and  $n_{B_i} (0)=1$ for $i = 1, \ldots, N$.
In a homogeneous setting where  firing delays of all workers are exponentially distributed with rate $\mu >0$, the steady-state forced idle probability 
	${\mathsf P}^{\ast}_{\rm idle}$ for each Worker$_i$ is given by 
\begin{equation}
	{\mathsf P}^\ast_{\rm idle} \;=\; \frac{N-1}{2N-1}.
    \label{eq:pblock}
\end{equation}
\end{proposition}

\begin{proof}[\textbf{Proof}]
 By the stars-and-bars theorem~\cite{graham1994concrete}, the number of ways to distribute $N$ indistinguishable tokens among $N$ distinguishable places is
\(
    \binom{2N-1}{N-1}
\). 
Worker$_i$ is idle if place $B_{i}$ contains no token. Fixing one place to be empty reduces the problem to distributing $N$ tokens among $N-1$ places, yielding
\(
    \binom{2N-2}{N-2}
\). 
Therefore, the steady-state forced idle probability is given by
\[
    {\mathsf P}^\ast_{\rm idle} 
    \;=\; \frac{N-1}{2N-1}. \qedhere
\]
\end{proof}
Given ${\mathsf P}^{\ast}_{\rm idle}$, a worker is
active for a fraction $1 - {\mathsf P}^{\ast}_{\rm idle}$ of the 
steady-state period.
Since an active worker completes firings at rate $\mu$, the
long-run effective firing rate with $N$ workers, also referred to as {\it throughput} and denoted by $\theta^* (N)$, is
given by  
\begin{align}\label{eq:mueff}
\theta^* (N) = \mu(1 - {\mathsf P}^\ast_{\rm idle}) \, . 
\end{align}
We can use the effective firing rate for a first-order estimate of the completion time~$T_{\rm RAR}$  of Ring AllReduce.  
Let $T_{{\rm RAR},i}^\ast$ denote the (random)  time of Worker$_i$ for completing $2(N-1)$ firings under steady-state assumptions. 
Given the effective throughput~$\theta^* (N)$ from Eq.~\eqref{eq:mueff}, an application of Little's Law~\cite{Little2011Reprint} yields the average 
\begin{align}\label{eq:Tss}
    \mathbb{E}[T_{\rm RAR}^{\ast}] \;=\;  \frac{2 (N-1)}{\theta^* (N)} = \frac{2 (N-1)}{\mu(1 - {\mathsf P}^{\ast}_{\rm idle})} \, .  
\end{align}

\begin{figure}[t!]
    \centering
   \includegraphics[width=0.7\linewidth]{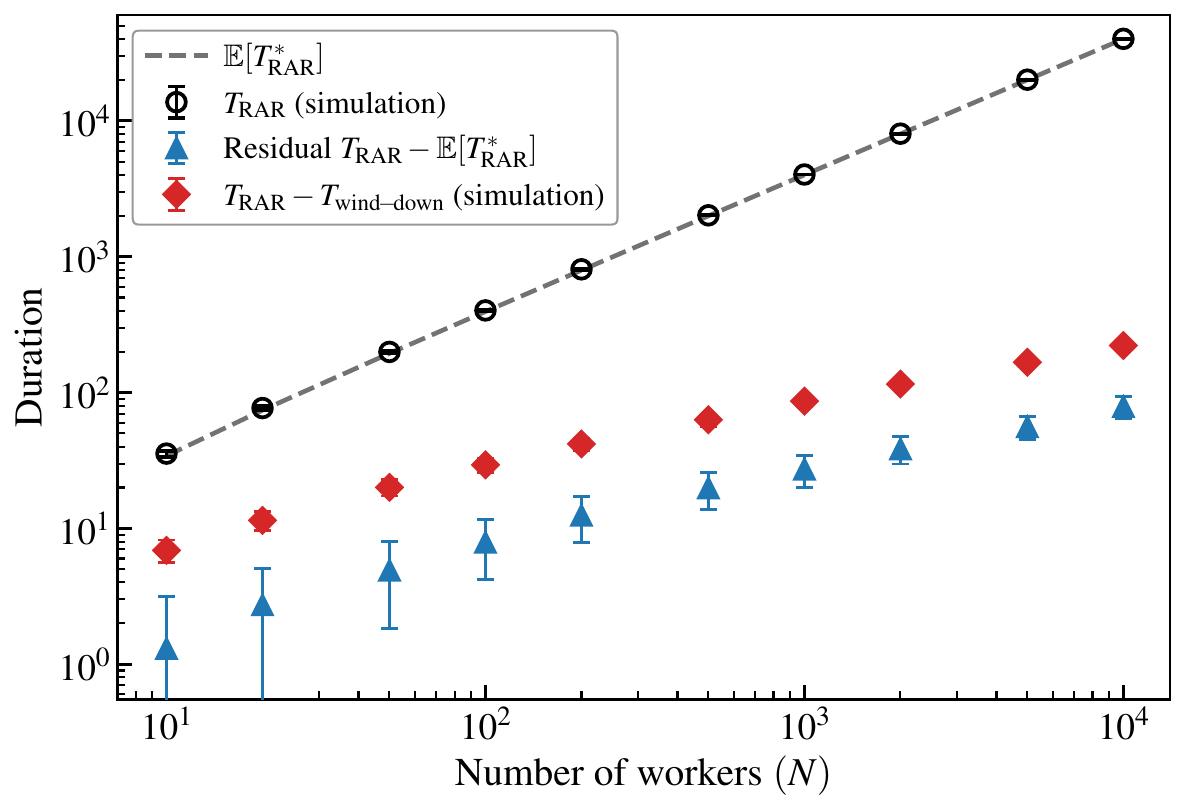}
    \caption{Completion times of Ring AllReduce.}
   \label{fig:completion_time}
\end{figure}

In Fig.~\ref{fig:completion_time}  we show that $\mathbb{E}[T_{\rm RAR}^{\ast}]$ provides a useful estimate for ~$T_{\rm RAR}$. 
The figure compares results from a steady-state analysis to simulations, where each simulation comprises 500 runs, 
on a logarithmic scale for  $N=10-10,000$ workers. 
For all simulation results, the markers indicate averages and 
error bars indicate one sample standard deviations. 
The black circles in the figure give the average completion times obtained from simulations. The error bars are  small 
enough that they fit inside the circles. The black dashed line shows $\mathbb{E}[T_{\rm RAR}^{*}]$ from Eq.~\eqref{eq:Tss}. 

Blue triangles and red diamonds, respectively, in the figure indicate the 
residual 
$T_{\rm RAR} - \mathbb{E}[T_{\rm RAR}^*]$ and the start of the 
wind-down phase $T_{\rm wind-down}$ 
over 500 simulations. 
The markers are the averages and the error bars represent 
one standard deviation. 
The figure illustrates that both the residual and the wind-down duration have sublinear scaling. The power law of this scaling will be made precise in  Sec.~\ref{sec:tasep}. 
Before exploring the scaling of the wind-down phase, we will investigate two variations of the SPN model. In Subsec.~\ref{subsec:Homo-Shift-Exp} we consider a system with limited randomness, where  firing delays have a constant component. In Subsec.~\ref{subsec:pfqn} we consider a system with stragglers, where some workers have longer firing delays.

\subsection{Shifted-Exponential Firing Delays}
\label{subsec:Homo-Shift-Exp}

We next consider that 
 computation and communication at workers in each 
phase of the Ring AllReduce operation incur a minimum delay. 
For this, we extend the homogeneous model by introducing a shifted-exponential
distribution of firing delays, which captures both a constant latency
component and a stochastic component.
Note that this extension no longer yields a Markov chain. 


The shifted-exponential firing delay Worker$_i$ is expressed as
a convex combination
\begin{equation}
    S_i \;=\; (1-c) X_i \;+\; \frac{c}{\mu},\qquad
    X_i \;\sim\; 
   \mathrm{Exp}(\mu) \, ,
    \label{eq:shifted_exp}
\end{equation}
where $c \in [0,1)$ controls the deterministic fraction,
and the exponential variables $X_i$ are independent.
With this choice, the shifted-exponential and pure-exponential 
models have the same average service time  $1/\mu$.
Selecting $c = 0$ creates the pure-exponential case. Increasing $c$ 
reduces the stochastic
component of the service time and makes the service time less variable.
For $c \to 1$, we recover the deterministic 
scenario with constant firing delay~$1/\mu$.

\begin{proposition}[Steady-state Idle Probability under Shifted-exponential Firing Delays]
\label{thm:shifted}
For the SPN model of Ring AllReduce from Fig.~\ref{fig:spn} with 
$N$~workers, with initial marking 
$n_{A_i} (0)=\infty$ and  $n_{B_i} (0)=1$ for $i = 1, \ldots, N$, consider 
a homogeneous scenario where each worker has shifted-exponential 
firing delays given by Eq.~\eqref{eq:shifted_exp}. 
Then the steady-state forced idle probability  of 
any worker is given by 
\begin{equation}
{\mathsf P}^\ast_{\rm idle} 
   \;=\; \frac{ (1-c) p^*}
{1-c p^*}\,,\quad \text{where}\ p^* = \frac{N-1}{2N-1}\,.
    \label{eq:pblock_shifted}
\end{equation}
\end{proposition}
Note that for $c=0$, Eq.~\eqref{eq:pblock_shifted} reduces to
$p^*=(N-1)/(2N-1)$, consistent with Lemma~\ref{lem:pblock}. 
As $c\to 1$ the system converges to the deterministic
model where workers are never idle.

\begin{proof}
We will reverse the argument in
Eqs.~\eqref{eq:pblock}--\eqref{eq:Tss} and first obtain an expression
for the expected time required for a worker to complete
$r$ firings. 
By taking $r\to\infty$, we deduce
the idle probability with Little's law.
As in the previous section, we start 
in the initial state $n_{B_i}(0)=1$.
As before, we denote the $r$-th firing delay of Worker$_i$ by 
$S_{i,r}$. 

\begin{enumerate}
\item {\em The stopping times $\tau_{i,r}$.}
For the shifted exponential model 
with firing times given by Eq.~\eqref{eq:shifted_exp},
let $\tau_{i,r}$ be the time when the $i$-th worker
completes its $r$-th firing, as defined in Eq.~\eqref{eq:tau-ir}.
Here, $i=1,\ldots, N$ and $r\ge 0$.
Denote the corresponding stopping times 
for the pure exponential model with $c=0$
by~$\tau'_{i,r}$.
We will prove by induction over $r$ that
\begin{equation}
\label{eq:shift}
\tau_{i,r} = (1-c)\tau'_{i,r} + c\frac{r}{\mu}
\end{equation}
for all $r\ge 0$ and $i=1,\ldots, N$.

{\em Base case.} For $r=0$ there is nothing
to show, since $\tau_{i,0}=0=\tau'_{i,0}$
by convention.

{\em Inductive step.} Assume as the inductive hypothesis
that Eq.~\eqref{eq:shift} holds for some $r\ge 0$,
for all $i=1,\ldots, N$.
Then, by Lemma~\ref{lemma:stopping_time},
\begin{align*}
\tau_{i,r+1} &= 
\max \left\{ \tau_{i,r}, \tau_{i-1,r}\right\}
+  S_{i,r+1}\\
&= \Bigl((1-c) \max \left\{ \tau'_{i,r}, \tau'_{i-1,r}\right\}
+ c\frac{r}{\mu} \Bigr)+ 
\Bigl((1-c) X_{i,r+1}
+  c\frac{1}{\mu}\Bigr)\\
&= (1-c) \tau'_{i,r+1} + c\frac{r+1}{\mu}\,,
\end{align*}
where the first equality is the renewal
relation in Eq.~\eqref{eq:renewal}
for the shifted-exponential system,
the second applies the inductive hypothesis and collects
terms, and the last uses the renewal relation for the pure
exponential. This completes the inductive step,
and establishes Eq.~\eqref{eq:shift}.

\item {\em Passing to the steady-state.} 
By definition, in the shifted-exponential model,
Worker$_i$ completes exactly $r$ firings
in the time interval $[0,\tau'_{i,r}]$.
By Little's law, the ratio $r/\tau_{i,r}$ converges
almost surely to the steady-state throughput, $\theta^*$.
It follows that
\begin{align*}
\frac{1}{\mu(1- {\mathsf P}^\ast_{\rm idle})} \ 
&= \ \lim_{r\to\infty} \frac{\mathbb{E}[\tau_{i,r}]}{r}\\
&= \ (1-c)\lim_{r\to\infty}\frac{\mathbb{E}[\tau'_{i,r}]}{r}
\ + \ \frac{c}{\mu}\\
&= \ \frac{1-c}{\mu(1- p^*)}\ +\ \frac{c}{\mu}\,.
\end{align*}
In the first line we have used that 
each worker operates at rate $\mu$ and used Little's law,
and in the second
line  we have applied Eq.~\eqref{eq:shift}.
In the last line we have applied Proposition~\ref{lem:pblock},
noting that $p^*=(N-1)/(2N-1)$.
The conclusion in Eq.~\eqref{eq:pblock_shifted}
is reached by solving
for ${\mathsf P}^\ast_{\rm idle}$.
\end{enumerate}
\end{proof}
Since $p^*\to \frac12$ for large $N$, the forced 
idle probability converges as well:

\begin{corollary}\label{coro:shifted}
In the SPN model with shifted-exponential firing delays given by 
Eq.~\eqref{eq:shifted_exp},  the forced idle probability converges to 
\[
   {\mathsf P}^\ast_{\rm idle} \underset{N \to \infty}{\longrightarrow}\frac{1-c}{2-c} \, . 
\]
\end{corollary}

\begin{figure}[t!]
\centering 
\subfloat[Forced idle probability ${\mathsf P}^\ast_{\rm idle}$.]{\includegraphics[width=0.49\textwidth]
{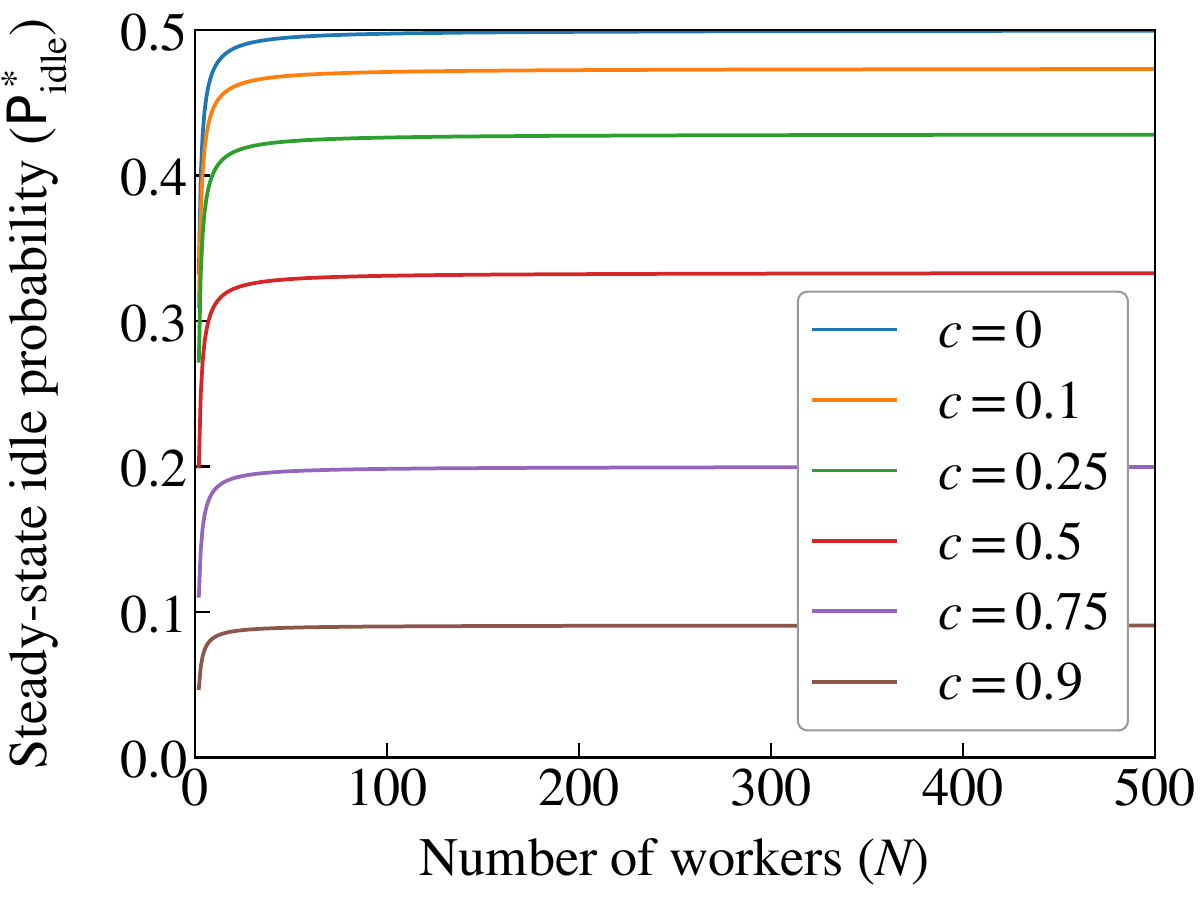}
 \label{fig:ss_idle_vs_N}}
\subfloat[Completion times.]{\includegraphics[width=0.49\textwidth]{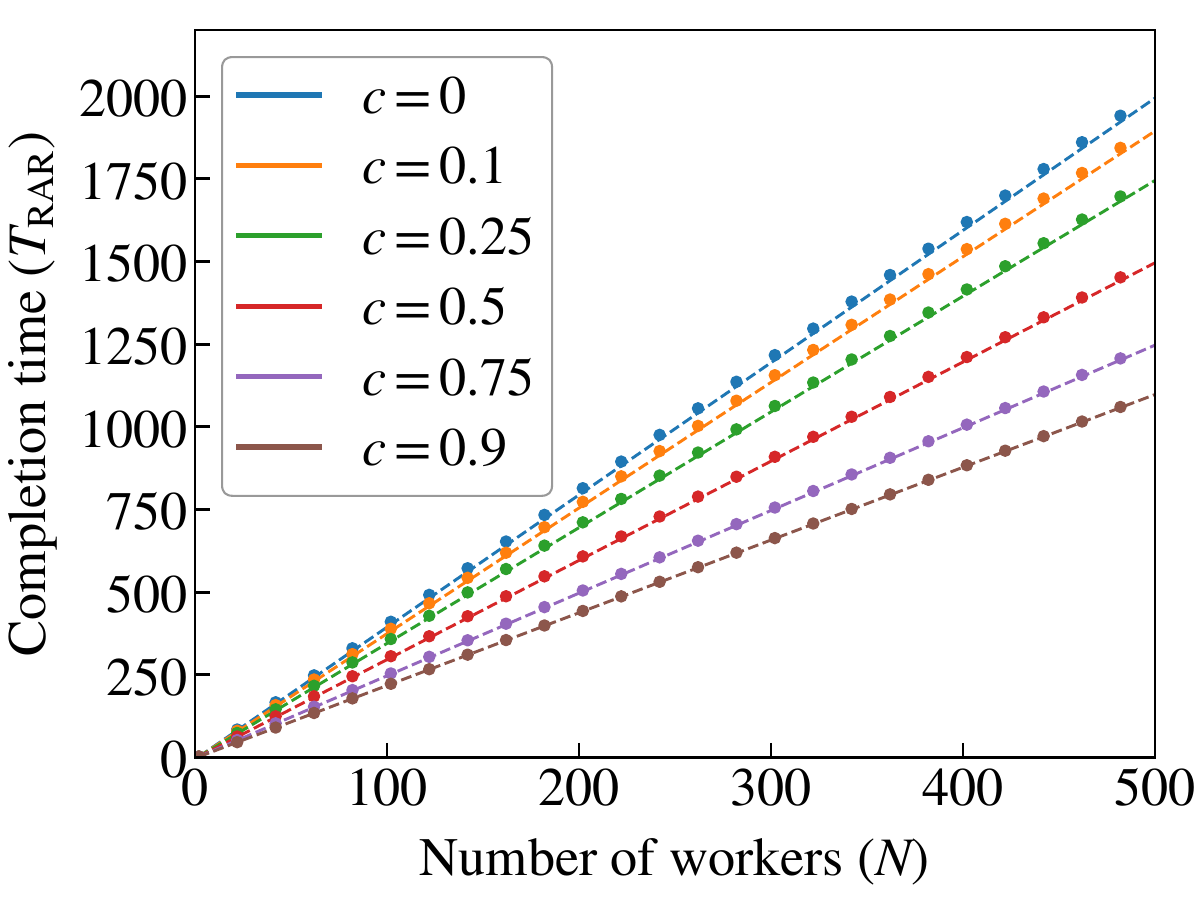}
\label{fig:ss_Tcomp_vs_N}}

\caption{Shifted-exponential firing delays with constant component $c/\mu$.} 
\label{fig:shifted-exp}
\end{figure}

Fig.~\ref{fig:shifted-exp} evaluates the force idle probability and completion times for 
the shifted-exponential scenario. We consider a system with up to 500 workers with 
rate $\mu=1$.  Fig.~\ref{fig:ss_idle_vs_N} shows the impact of the constant component 
on the idle probability ${\mathsf P}^\ast_{\rm idle}$ from Eq.~\eqref{eq:pblock_shifted}. Note that 
for all values of $c$, the asymptotic  value given by Corollary~\ref{coro:shifted}
is approached quickly when increasing the number of workers.   The forced idle probability decreases as the deterministic 
component is increased and is zero for a purely deterministic system. In Fig.~\ref{fig:ss_Tcomp_vs_N} we evaluate  completion times, where markers indicate simulations of $T_{\rm RAR}$  and lines are  the steady-state expected completion time 
$\mathbb{E}[T_{\rm RAR}^{*}]$ from Eq.~\eqref{eq:Tss}.  The figures indicate 
that reducing the variable (random) component has a significant impact on completion times.

\subsection{Heterogeneous Firing Delays (Stragglers)}
\label{subsec:pfqn}

A well-known obstacle in distributed training is the presence of {\it stragglers}, that is, workers that incur relatively long computation or communication delays and thereby slow down overall progress.
For considering stragglers in the SPN model of Ring AllReduce, we need to account for  
transitions with heterogeneous firing delays. When service rates are
heterogeneous, the stars-and-bars argument for closed-form expressions of the forced idle
probability no longer applies. 
To enable an analytical evaluation under heterogeneous firing delays, we map the Ring AllReduce SPN model to a closed queueing network. As before, we conduct a steady-state analysis of the SPN model from Fig.~\ref{fig:spn}, where we ensure a perpetual continuation of operations on the ring by eliminating places~$A_i$. Then, for each Worker$_i$, we replace place $B_i$ and transition~$T_i$ by a single-server FCFS (First-Come-First-Served) queue 
 with exponential service times at rate $\mu_i$. 
Tokens of the SPN correspond to customers in the queueing network. (We continue to use the term `token'. )
By Lemma~\ref{Sum_N_b},
the total number of tokens across the queues satisfies
$\sum_{i=1}^{N} n_{B_i}(t) = N$ at all times $t$. Since the lemma 
holds for every $t \geq 0$, we omit the time index $t$ in the remainder of this 
section. 
Upon completing service, a token at the queue of Worker$_i$ is always
forwarded to that of Worker$_{i+1}$, hence, the routing process is irreducible, 
with visit ratio $e_i = 1$ for each queue~$i$. 
Without external arrivals or departures, the model is a closed queueing network that satisfies the conditions for a product-form solution~\cite{gordon1967closed,bcmp}. 
The state space of the queueing network is given by 
$\mathbf{n} = \left(n_{B_1},\ldots,n_{B_N}\right)$ with $n_{B_i} \geq 0$,
$\sum_{i=1}^{N} n_{B_i} = N$. Since the state space, the reachable states, and  transition rates match those of the modified SPN model from Eq.~\eqref{eq:state-spn}, 
their underlying Markov chains are equivalent. 
Hence, we can analyze the SPN model as a product-form queueing network. 
In the following we will use `Worker$_i$' to refer to the queue for Worker$_i$. 

From~\cite{gordon1967closed},  
the steady-state probability of state  $\mathbf{n}$, denoted by $\pi(\mathbf{n})$, takes the product form
\begin{equation}
    \pi(\mathbf{n}) \;=\; \frac{1}{G_N}
    \prod_{i=1}^{N} \frac{1}{\mu_i^{n_{B_i}}} \, ,
    \label{eq:product_form}
\end{equation}
where $G_N$ is the normalization constant with population $N$, which ensures that the sum of the probabilities of all possible states equals one. 
For our purposes, it is sufficient to analyze the network via the Mean Value Analysis
(MVA)~\cite{reiser1980mean}, thereby avoiding the computation of the normalization constant or an enumeration of the full state space.


Let ${W}^*_i(k)$ denote the mean sojourn time of a token at Worker$_i$, including both queueing delay and service time, when the total
number of circulating tokens in the ring is $k$. Let ${Q}^*_i(k)$
denote the corresponding mean number of tokens at Worker$_i$. 
The iterative MVA algorithm is initialized with ${Q}^*_i(0)=0$ for all $i$ and proceeds for $k=1, 2, \ldots$ as 
\begin{align}
    {W}^*_i(k) &\;=\; \frac{1}{\mu_i}
    \bigl(1 + {Q}^*_i(k-1)\bigr)\, ,
    \label{eq:mva_W} \\
    \theta^*(k) &\;=\; k \left(\sum_{i=1}^{N}
    {W}^*_i(k) \right)^{-1}\, ,
    \label{eq:mva_theta} \\
    {Q}^*_i(k) &\;=\; \theta(k)\,{W}^*_i(k) \, ,
    \label{eq:mva_Q}
\end{align}
where $\theta^*(k)$ is the throughput with population $k$. 
The iteration ends with the total population $k=N$.  
The throughput of the queueing 
network corresponds to the throughput obtained for the SPN as given in Eq.~\eqref{eq:mueff}. 
Due to the circular routing,  all queues have the same throughput  value $\theta^*(N)$. 
The fraction of time that the queue for Worker$_i$ contains at least one token, referred to as utilization, is 
given by   $\theta^*(N)/ \mu_i$.  Hence, the steady-state probability 
${\mathsf P}^\ast_{{\rm idle},i}$ that Worker$_i$ holds no token  is  
\begin{equation}
    {\mathsf P}^\ast_{{\rm idle},i}  \;=\; 1 \;-\; \frac{\theta^*(N)}{\mu_i} \, . 
    \label{eq:mva_pss}
\end{equation}
This probability corresponds to the forced idle probability in the SPN given 
in Eq.~\eqref{eq:mueff}. 
In the homogeneous case, where $\mu_i=\mu$ for all $i$, 
we can solve for $\theta^*$ in Eq.~\eqref{eq:mva_pss}
to recover the closed-form expression in 
Proposition~\ref{lem:pblock}.

As in the homogeneous case (see Eq.~\eqref{eq:Tss}), we can use Little's Law to express 
the expected completion time of Worker$_i$ under steady-state assumptions as 
\begin{align}\label{eq:Tss-het}
    \mathbb{E}[T_{{\rm RAR}}^{\ast}] \;=\;  \frac{2 (N-1)}{\theta^* (N)} \, .  
\end{align}
As in Eq.~\eqref{eq:Tss}, since all workers have the same expected completion time, we drop the subscript from $T_{{\rm RAR},i}^{\ast}$.

\begin{figure}[t!]
\centering
\subfloat[Idle probability.]{\includegraphics[width=0.5\textwidth]{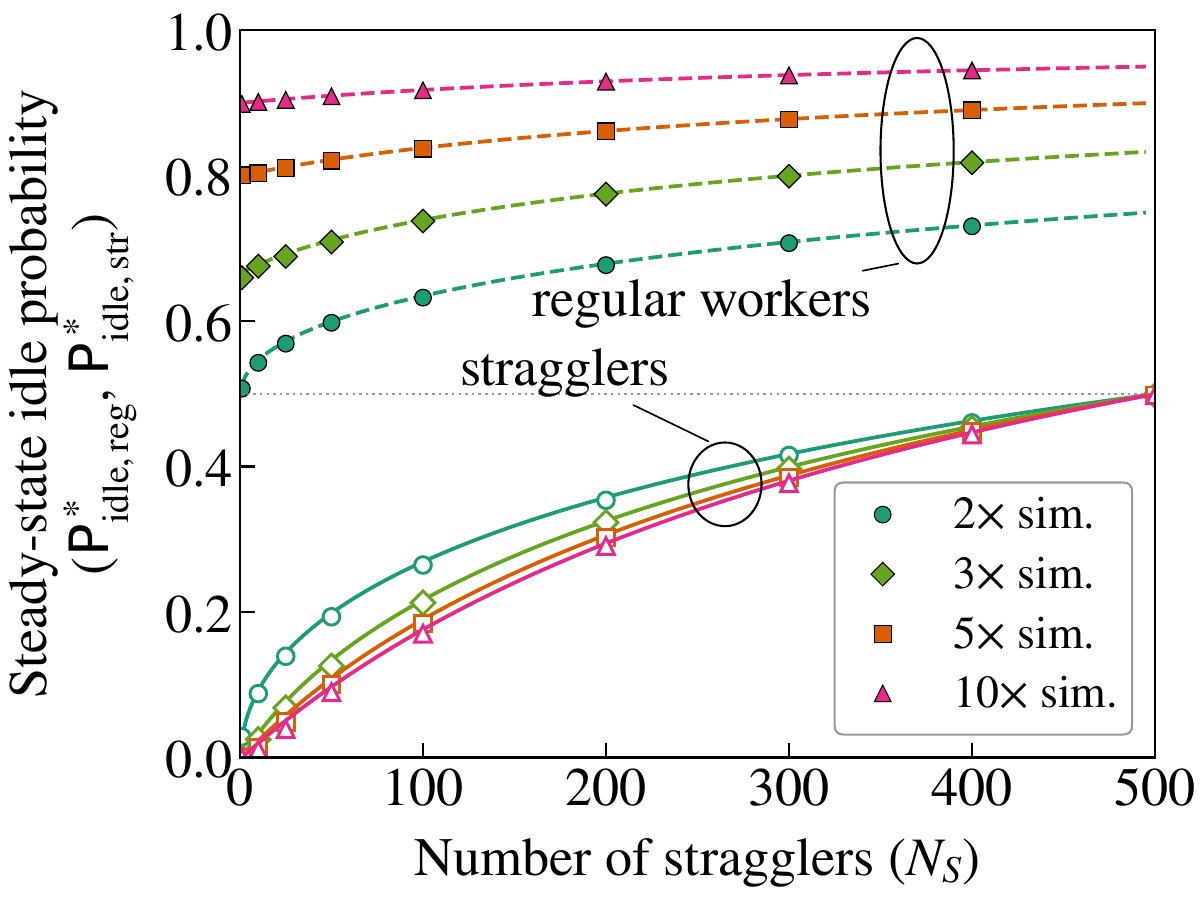}\label{fig:mva-idle}}
\subfloat[Throughput.]{\includegraphics[width=0.5\textwidth]{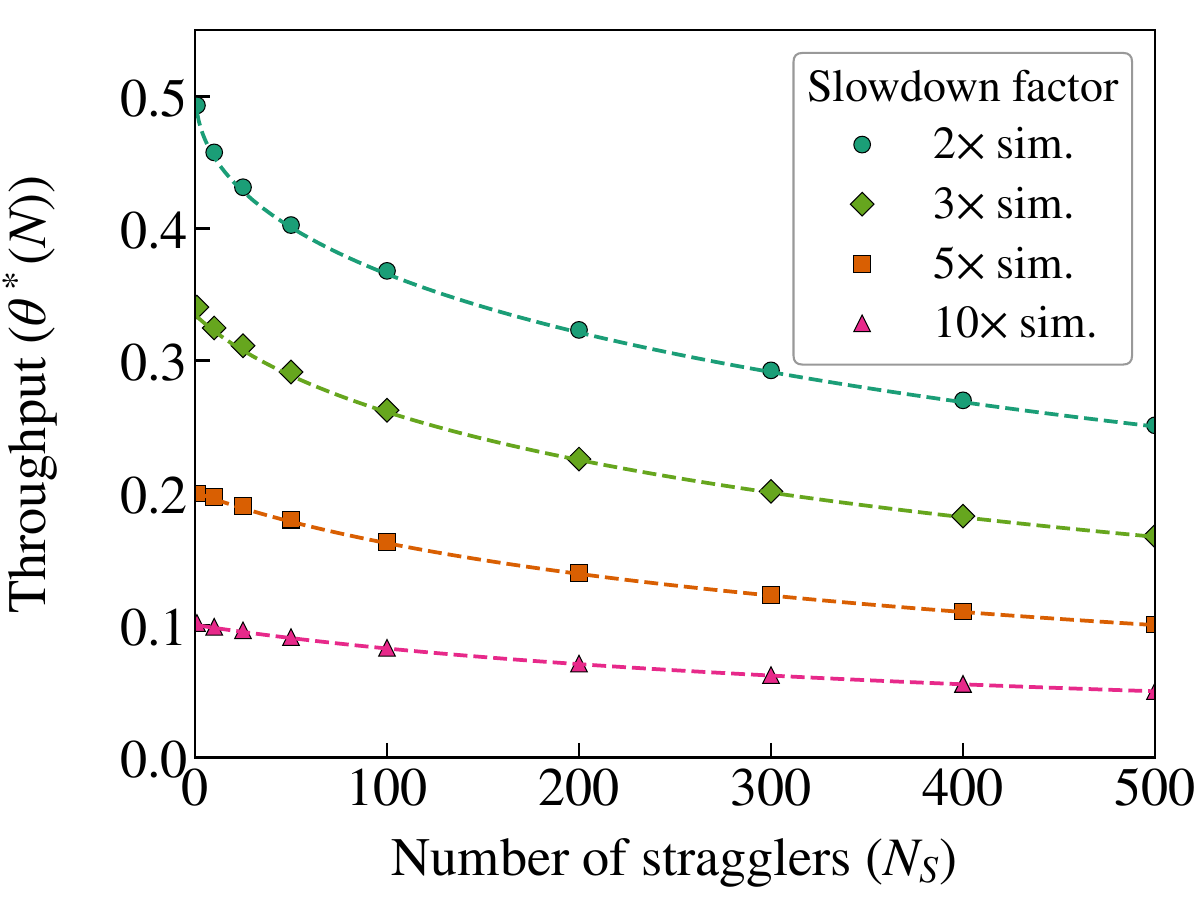}\label{fig:mva-tput}}   

\centering 
\subfloat[Average number of idle workers.]{\includegraphics[width=0.5\textwidth]{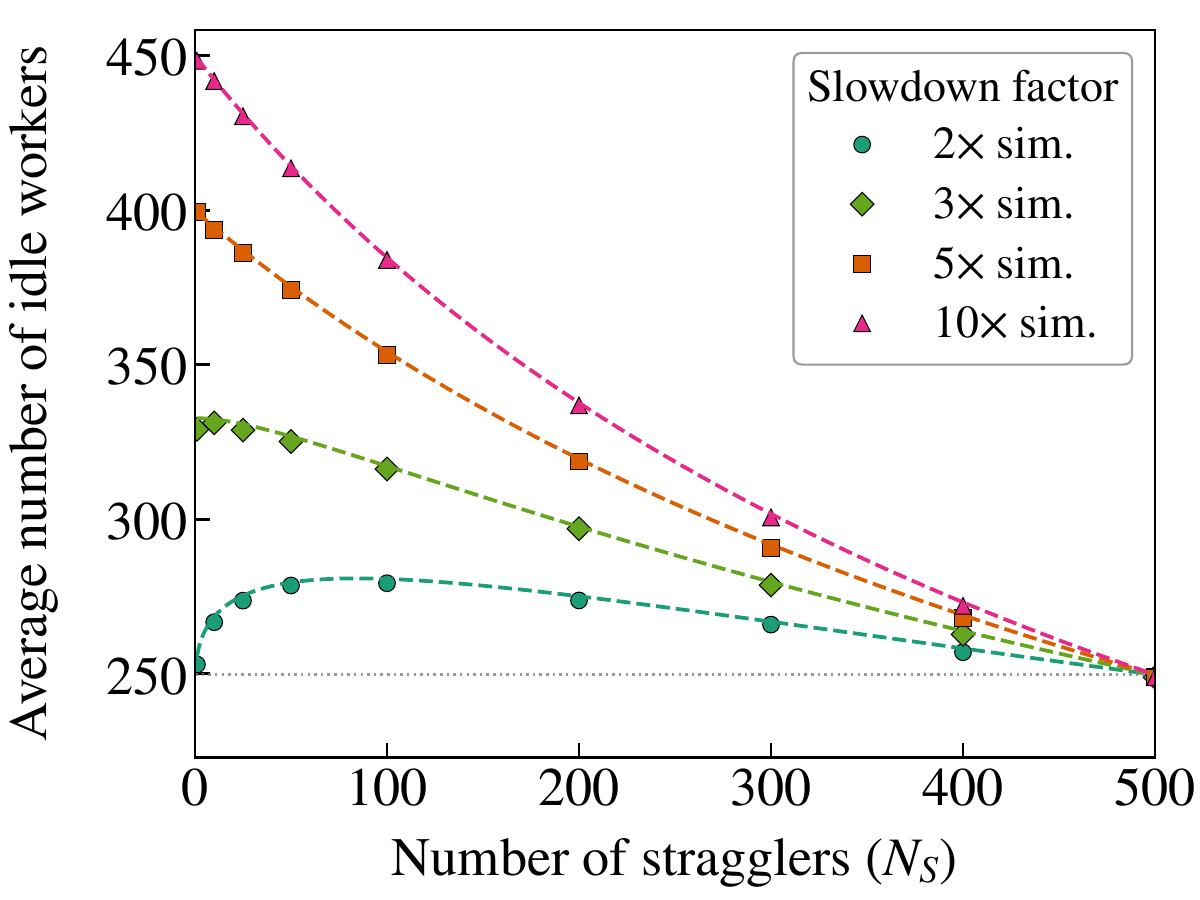}\label{fig:mva-idle-avg}} 
        \caption{Mean value analysis of a Ring AllReduce model with $N=500$ workers with stragglers. MVA analysis of a steady-state system is shown as solid or dashed lines. Simulations of Ring AllReduce with $2 (N-1)$ steps are indicated by markers. }
    \label{fig:mva_stragglers} 
\end{figure}


\medskip 
Let us consider a scenario with $N=500$ workers,
where $N_S$ workers are stragglers with service rate
$\mu/s$  ($s>1$) and $N-N_S$ regular workers with service rate 
$\mu$, with $\mu=1$. 
We refer to $s$ as {\it slowdown factor}. 
Note that, with a  product-form queuing network, the positions of 
the stragglers in the ring have no impact on idle probabilities or throughput. 
The MVA estimate is obtained from the closed-network recurrence from 
Eqs.~\eqref{eq:mva_W}--\eqref{eq:mva_Q}.  
We denote the idle probabilities, 
obtained with Eq.~\eqref{eq:mva_pss}, of regular workers and stragglers 
by $ {\mathsf P}^*_{\rm idle, reg}$  and $ {\mathsf P}^*_{\rm idle, str}$, respectively.

For comparison, we also provide simulations of the Petri Net model of Ring AllReduce from Fig.~\ref{fig:spn} with initial 
markings $n_{A_i} (0)= 2(N{-}1)$. Our simulations measure the throughput 
and forced idle probability during the stable phase, 
where we  identify the end of the ramp-up phase and the start of the 
wind-down phase as discussed in Sec.~\ref{subsec:CTMC}. 
Simulations are repeated 20 times, and the depicted values are the averages of all runs. 

Fig.~\ref{fig:mva_stragglers} presents the idle probability, throughput, and 
average number of idle workers  as a function of the number of stragglers for 
slowdown factors $s = 2, 3, 5, 10$. 
In the figures, results from the mean value analysis are shown as solid or dashed line 
and simulations are indicated by markers. 
As a first observation, note that the mean value analysis closely 
matches outcomes of the (finite-term) simulations. 

Fig.~\ref{fig:mva-idle} plots the steady-state idle probabilities ${\mathsf P}^*_{\rm idle, str}$ and 
${\mathsf P}^*_{\rm idle, str}$ as solid and dashed lines, respectively. If the number of stragglers 
is small, the stragglers are busy most of the time. As the number of stragglers is increased, the idle 
probability ${\mathsf P}^*_{\rm idle, str}$ converges to  Eq.~\eqref{eq:pblock}, independent of the 
speed of the stragglers. 
This is different for the  regular (non-straggler) workers. Here, the idle probability 
${\mathsf P}^*_{\rm idle, reg}$ 
increases with the number of stragglers, yet, the idle probability continues to depend on the slow-down factor. 

Fig.~\ref{fig:mva-tput} shows that the speed of the stragglers 
has a large impact on the throughput. For $s=10$, even a single 
straggler drops the throughput to about 0.1, compared to a throughput of close to 0.5 (due to Eq.~\eqref{eq:pblock}). In contrast, 
for a slow-down factor $s=2$, the throughput is robust to the presence of a small number of stragglers.

In Fig.~\ref{fig:mva-idle-avg} we present the average number of idle workers, $N^*_{\rm idle}$ computed  from 
\[
 N^*_{\rm idle} = {\mathsf P}^*_{\rm idle, str}  \cdot N_S +  {\mathsf P}^*_{\rm idle, reg} \cdot (N-N_S)  \, . 
\]
Interestingly, even though both ${\mathsf P}^*_{\rm idle, str}$ and ${\mathsf P}^*_{\rm idle, reg}$ increase with the number of stragglers, 
the average number of idle nodes largely decreases (at the cost of a reduced throughput). The exception is the case $s=2$ where the number of
idle workers initially increases.

%% file: 5A_Mapping_to_TASEP_Nov25.tex
\section{TASEP Analysis of Ring AllReduce}
\label{sec:tasep}

In the previous section
we derived the steady-state forced idle probability
for homogeneous and heterogeneous service 
rates, using stars-and-bars and queueing network analysis. 
In this section, we refine our analysis
by considering the contribution of the 
wind-down phase to the
Ring AllReduce completion time. 
Our analysis also
addresses the systematic deviation of the
completion time $T_{{\rm RAR}}$
from the linear approximation in Eq.~\eqref{eq:Tss}, 
and the statistical fluctuations around the mean.

We consider the homogeneous SPN from 
Sec.~\ref{sec:spn-model} where all workers have
identical exponential firing rates $\mu=1$.
We proceed by coupling the SPN 
to a special case of the totally asymmetric exclusion
process (TASEP)~\cite{TASEP1,TASEP2}, a particle process that 
models directed phenomena such as the
propagation of a flame front across a plane. 
The TASEP model has received much attention in Probability and
Statistical Physics over the last 20 years~\cite{tracy-Widom2008tasep,BFPS_2007tasep,Quastel26}.
A hallmark of TASEP is that for certain key
random variables whose mean grows linearly in time,
typical fluctuations from the mean grow only
as $t^{1/3}$. This anomalous scaling stands in
sharp contrast with the central limit theorem for sums of i.i.d. 
random variables.

Consider the SPN model from Fig.~\ref{fig:spn}, where 
the initial number of tokens in places~$A_i$ ($i=1,\ldots,N$) 
is given by $n_{A_i}(0) = 2(N-1)$. With identical workers, 
TASEP theory predicts that the time when Worker$_i$ completes
its work deviates from $T^\ast_{\rm RAR}$
by a random error of order $N^{1/3}$, that is,
\begin{equation}
\label{eq:TASEP-i}
T_{{\rm RAR},i} \sim T^\ast_{\rm RAR} + N^{1/3}\xi
\qquad \text{as}\ N\to\infty\,.
\end{equation}
Here, $\xi$ is a random variable that does not depend on $N$, and
$\sim$ indicates that the random variables
\begin{equation}
\label{eq:xi-i}
\xi_i:=\left(T_{{\rm RAR},i} -T^\ast_{\rm RAR}\right)/N^{1/3}
\end{equation}
converge in distribution to $\xi$ as $N\to\infty$.
This limit will be discussed in more detail below. 
The same scaling will be applied to the
completion time
$T_{\rm RAR}$ and the start of the wind-down 
phase $T_{\rm wind-down}$,
\begin{align}
\label{eq:xi-RAR}
\left(T_{\rm RAR}-T^\ast_{\rm RAR}\right)/N^{1/3}\ &=\ \max_i \xi_i\,,\\
\label{eq:xi-wd}
\left(T_{\rm wind-down}-T^\ast_{\rm RAR}\right)/N^{1/3}\ &=\ \min_i \xi_i\,.
\end{align}

\begin{figure}[t!]
\centering 

\subfloat[Expected value $\mathbb{E}{[}\xi{]}$.]
{\includegraphics[width=0.48\textwidth]{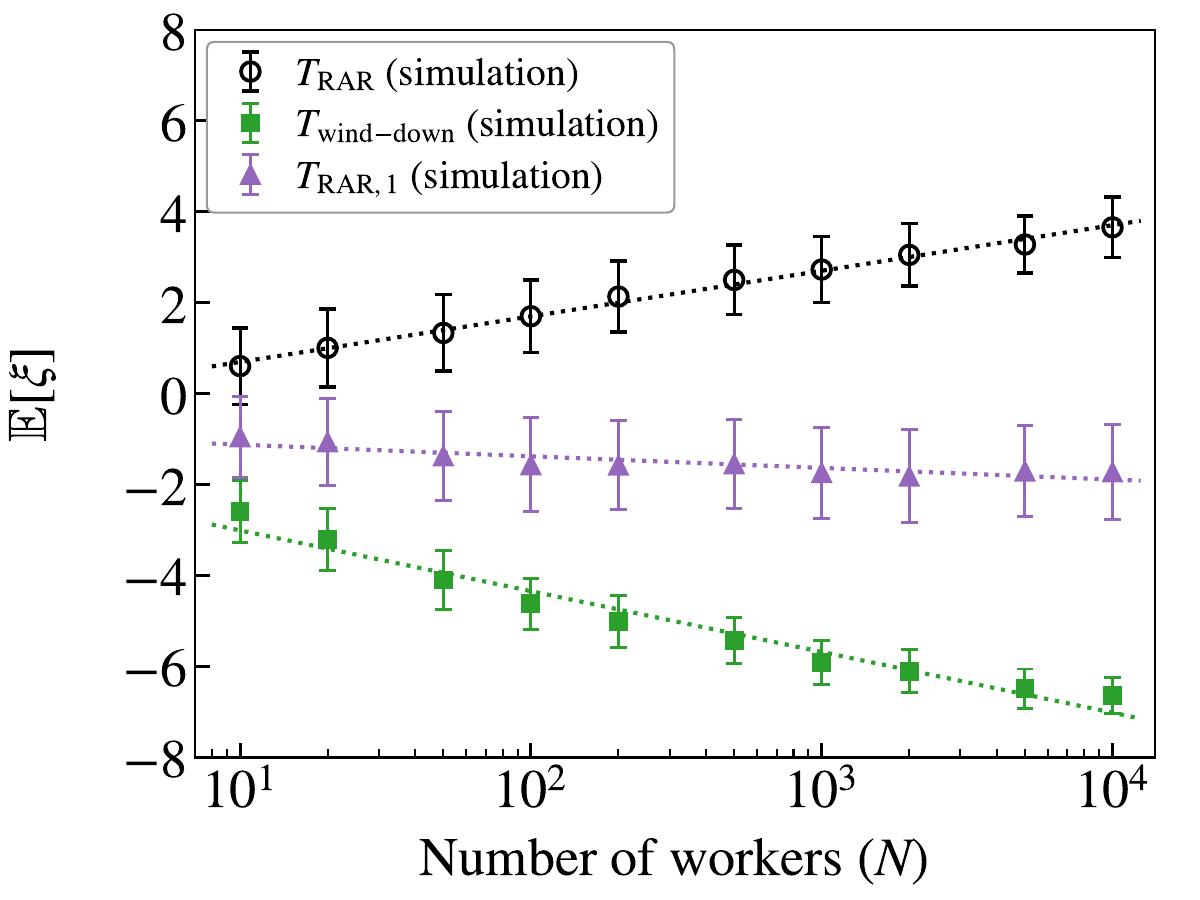}
\label{fig:wind_down_scaled}}
\subfloat[Standard deviation $SD(\xi)$.]
{\includegraphics[width=0.48\textwidth]
{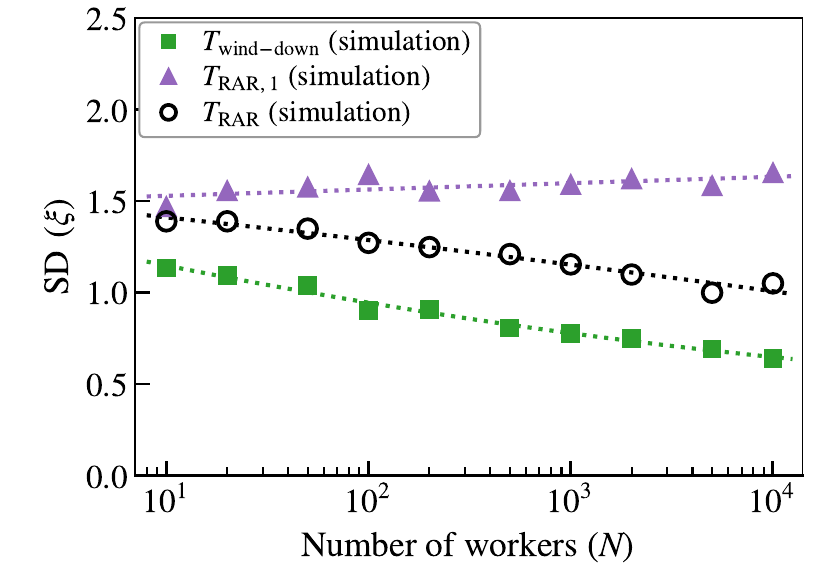}
 \label{fig:TASEP_scaled_statistical_fluctuations}}
	\caption{Fits of completion times with TASEP predictions. } 
\label{fig:shifted-exp}
\end{figure}

In Fig.~\ref{fig:shifted-exp} we present the fit of  
completion times for Ring AllReduce obtained from simulations 
to the TASEP model for 10 to 10,000 workers. 
The figure tracks three metrics: 
$T_{{\rm RAR},i}$ from Eq.~\eqref{eq:T-rar-i} with $i=1$,
$T_{\rm RAR}$ from Eq.~\eqref{eq:T-rar}, 
and $T_{\rm wind-down}$ from Eq.~\eqref{eq:T-wd},
which correspond to the completion
time of an individual worker in isolation,
the total completion time of Ring AllReduce the SPN model,
and the start of the wind-down phase.
Figs.~\ref{fig:wind_down_scaled}  
and~\ref{fig:TASEP_scaled_statistical_fluctuations}, 
respectively, present the 
expected values and standard deviations
of the scaled random variables $\xi_i$, $\max_i \xi_i$, and $\min_i\xi_i$
in Eqs.~\eqref{eq:xi-i}-\eqref{eq:xi-wd} as functions
of $N$, where the x-axis is plotted on a log scale. 
The markers in the graph are averages of 500 simulations, 
and the error bars indicate 
one sample standard deviation. 
Dotted lines in the plots are linear trend lines for the averages. 
We observe that for a worker in isolation,
$\mathbb{E}[\xi]$ and $\mathrm{SD}(\xi)$ appear 
essentially constant, confirming that 
the completion time of an individual worker is well described 
by Eq.~\eqref{eq:TASEP-i}. 

In contrast, the other two curves in Fig.~\ref{fig:wind_down_scaled}
show diverging linear trends for 
$T_{\rm RAR}$ and $T_{\rm wind-down}$.
This suggests that each requires a logarithmic correction 
to the TASEP model. 
We will argue that the corresponding
scaled variables are well-approximated by
\begin{align}
\label{eq:TASEP-RAR}
\max_i\xi_i \ &\approx C_{\rm RAR}\log (N) + \xi_{\rm RAR},
\\
\label{eq:TASEP-wd}
\min_i\xi_i \ &\approx C_{\rm wind-down}\log (N) + \xi_{\rm wind-down},
\end{align}
where $C_{\rm RAR}>0>C_{\rm wind-down}$ are constants,
and $\xi_{\rm RAR}$ and $\xi_{\rm wind-down}$
are random variables whose distributions do not depend on $N$.

The remainder of the section is organized as follows. 
In Subsec.~\ref{subsec:tasep_def} we define TASEP 
on a ring structure. This process is mapped to our SPN 
model for Ring AllReduce in Subsec.~\ref{subsec:mapping}. 
Then, in Subsec.~\ref{subsec-predictions}, we prove
Eq.~\eqref{eq:TASEP-i} and justify
the logarithmic correction 
for $T_{\rm RAR}$ and $T_{\rm wind-down}$.

\subsection{TASEP: Definition and Ring Structure}
\label{subsec:tasep_def}

\begin{figure}[t!]
\centering

\subfloat[Before transition.]
{\includegraphics[width=0.4\textwidth]{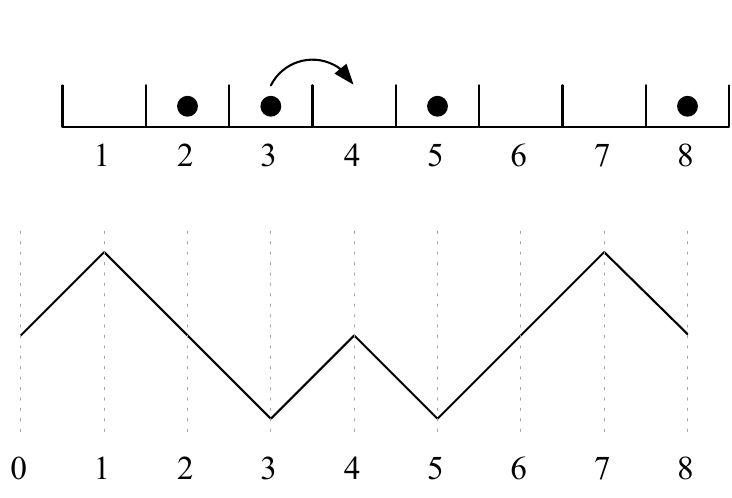}
\label{fig:before}}
\quad
\subfloat[After transition.]
{\includegraphics[width=0.4\textwidth] {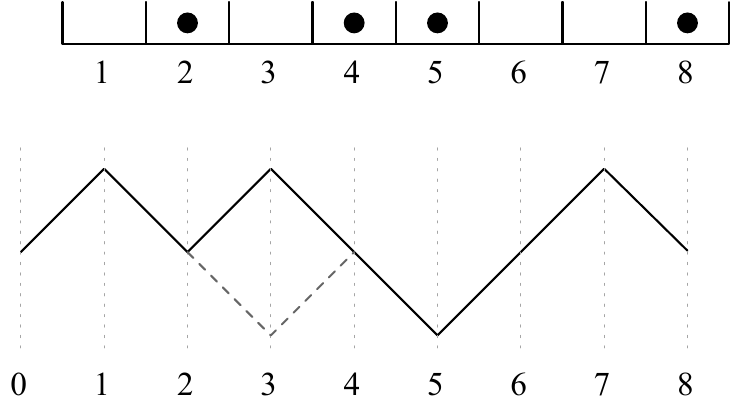}
 \label{fig:after}}
	\caption{Transition in TASEP on a ring with eight positions (The right-hand neighbor of 
	position~8 is position~1). Top graphs indicate
occupancy of positions; bottom graphs show the height function.
}
\label{fig:hop}
\end{figure}

TASEP is a CMTC that models the
dynamics of a collection of particles 
positioned on the integer line or on a ring.
Every state transition in TASEP describes the 
movement of a particle   to a neighboring position, subject to
the following conditions:
\begin{enumerate}
\item {\em Simple exclusion:} every position contains at most
one particle.
\item{\em Total asymmetry:} only hops to the right are allowed.
\end{enumerate}
We consider TASEP on a ring at half-filling, that is, 
where~$N$ particles are placed on a ring of size $2N$. 
The state of the system is given
by a binary string $\eta=(\eta_1,\ldots, \eta_{2N})$
that contains exactly $N$ ones,
and where the index is interpreted modulo $2N$.
We say that position $x\in \{1,\ldots, 2N\}$ 
is {\em occupied} by a particle
if $\eta_x=1$, and {\em vacant} if $\eta_x=0$.

Denote by $\eta(t)$ the state at time $t\ge 0$.
The possible transitions out of $\eta(t^-)$
correspond to positions $y$ 
with $\eta_y(t^-)=1$ and $\eta_{y+1}(t^-)=0$.
After transition, the state becomes
\begin{align}
\label{eq:hop}
\begin{split}
    \eta_{y}(t)     &\;\longleftarrow\; 0, \\
    \eta_{y+1}(t)     &\;\longleftarrow\; 1, 
\end{split}
\end{align}
with $\eta_x(t)=\eta_x(t^-)$ for all $x \neq y, y+1$
Such a transition is called a {\em hop}
of the particle at $y$ to its right neighbor at $y+1$. 
The TASEP model assumes that
all transitions occur at the same exponential rate $\mu=1$. Clearly, 
hops preserve the total number of particles $N=\sum_{x=1}^{2N} \eta_x$. 

Fig.~\ref{fig:hop} illustrates a transition in a TASEP on a ring where 
four particles (black circles) are placed on eight 
positions. In the figures, the ring is unrolled, that 
is, the right neighbor of position 8 is position 1.  
The top graphs in Fig.~\ref{fig:before} and~\ref{fig:after} show the occupancy 
before and after a hop at $y=3$.  The states before and after the transition at time $t$ 
are given by $\eta(t^-) =(0, 1, 1, 0,  1, 0, 0, 1)$ and $\eta (t)=(0, 1, 0, 1,  1, 0, 0, 1)$. 
 
The evolution of a TASEP is represented 
by a {\it height function}. 
The height function, $x\mapsto h_x$
is a nonnegative integer-valued function
with the
property that $|h_x-h_{x-1}|=1$ for all $x$, 
where in a ring, the index $x$ is interpreted modulo $2N$.

It is determined, up to an additive constant, by the state $\eta$ 
through the recursion 
\begin{align}
\label{eq:h-diff}
h_x= h_{x-1}+ 1-2 \eta_x\,, \quad x= 1, \ldots, 2N\,.
\end{align}
When moving in the clockwise direction
along the ring, each vacant position 
contributes $+1$ to the height
function, and each occupied position contributes $-1$.

In the transition given in Eq.~\eqref{eq:hop}, the height function is updated by
\begin{align}
\label{eq:hop-hy}
    h_{y}(t)     \;\longleftarrow\; h_{y}(t^-) +2\,,
\end{align}
while its values at all other positions remain unchanged. Since 
each transition increases the height function in one position, 
its values track the progress of the TASEP. 
The bottom graphs in Fig.~\ref{fig:hop} depict the height functions  
before and after the movement of a particle from position~$3$ to position~$4$. 
To illustrate the height function defined on a ring, we include a position 0 
to represent position 8. Note that our definition of the height 
function requires half-filling, since the number of increments 
(for empty positions) must match the number of 
decrements (for occupied positions). 

\subsection{Mapping TASEP on a ring  to the SPN model of Ring AllReduce}
\label{subsec:mapping}

We next construct the state space mapping from
TASEP on $\{1,\ldots, 2N\}$ to the Ring AllReduce SPN with $N$ workers.
The mapping interprets the TASEP occupancy vector
$\eta$ as a stars-and-bars representation
of the SPN markings $(n_{B_1},\ldots, n_{B_N})$,
anchored at a tagged vacant position $x_0$.
Given a TASEP state $\eta$ and a tagged vacant site
$x_0$, let $x_i$ be the $i$-th
vacant position in the clockwise direction from $x_0$. Since
there are exactly $N$ vacant positions in the ring,
$x_N=x_0$, and we may consider the index $i$ modulo $N$.
Moreover, $\sum_{i=1}^N (x_i-x_{i-1}-1)=N$.

The transition in Eq.~\eqref{eq:hop} moves one of the vacant 
positions to the left, that is, 
\begin{align}
\label{eq:hop-xi}
    x_i(t)     \;\longleftarrow\; y \qquad \text{if}\ x_i(t^-)=y+1,
\end{align}
while all other vacant positions  remain unchanged. 
The height function
satisfies $h_y(t)=h_y(t^-) =h_{y+1}(t^-)+1$,
since the position at $y$ is occupied at time $t^-$,
and therefore
\begin{align}
\label{eq:hop-hi}
    h_{x_i(t)}(t) \; &\longleftarrow\; h_{x_i(t^-)}(t^-)+1 \,.
\end{align}

Consider first the simplified SPN from 
Sec.~\ref{sec:idle_time_analysis}, where we take
the initial markings $n_{A_i}(0)\to\infty$ for
$i=1,\ldots, N$.
The state is given by ${n}=(n_{B_1},\ldots, n_{B_N})$ as in Eq.~\eqref{eq:state-spn}.
Define the mapping $\Phi: (\eta,x_0) \mapsto 
\mathbf{n}$ by 
$$
n_{B_i} := x_i-x_{i-1}-1\,,\qquad \text{for}\ i=1,\ldots, N\,,
$$
that is, the number of tokens held by Worker$_i$
equals the number of particles between
$x_{i-1}$ and $x_i$. 
The mapping $\Phi$ is surjective but cannot be
injective, since a rotation of the position labels in the ring
does not change the value of $\Phi(\eta,x_0)$.

\begin{proposition}\label{prop:Phi}
Let $\eta(t)$ be the occupation vector
of a TASEP with $N$ particles on a ring of length $2N$
at $t$, and let $x_0(t)$ be a 
tagged vacant position. Assume that the initial state is 
\begin{align*}
\eta_x(0)= \begin{cases} 0, \quad &\text{if $x$ is even,}\\
1, \quad &\text{if $x$ is odd,}
\end{cases}
\end{align*}
and set $x_0(0)=2N$.
Then the process 
$\Phi (\eta(t), x_0(t))$  
agrees with the simplified SPN from
Sec.~\ref{sec:idle_time_analysis}.
\end{proposition}

\begin{proof} 
In the TASEP initial state, the vacant
positions are $x_i(0)=2i$ for $i=1,\ldots, N$.
By definition of $\Phi$ this yields the
initial markings $n_{B_i}(0)=1$ for $i=1,\ldots, N$.

Under the mapping $\Phi$,
a hop of a particle at $y$ to an empty 
position at $x_i=y+1$, described by
Eqs.~\eqref{eq:hop} and~\eqref{eq:hop-xi},
results in a firing of Worker$_i$
that decrements $n_{B_i}$ and increments $n_{B_{i+1}}$,
in agreement with Eq.~\eqref{eq:MC-update}.
In the top parts of Figs.~\ref{fig:before} 
and~\ref{fig:after}, this is illustrated 
for the hop where the particle at position
$y=3$ changes place with
the vacant site at $x_i=4$.

All transitions occur at same exponential rate in the TASEP and 
in the simplified SPN.
\end{proof}

The relationship of TASEP with the
AllReduce SPN described in Sec.~\ref{sec:spn-model}
is more complicated, because the TASEP continues indefinitely,
while the SPN reaches an absorbing state.
To define a mapping from TASEP to
the SPN, we describe the TASEP state by the height function, $h$,
together with a tagged position $x_0$.
The occupation vector $\eta$ can be recovered
from the height function by setting
$$
\eta_x = \frac12(h_{x-1}-h_x+1)\,.
$$
To ensure that the tagged position is vacant,
we require that $h_x-h_{x-1}=1$.

Define $\Psi: (h,x_0) 
\mapsto \bigl(n_{A_1},n_{B_1},\ldots, n_{A_N}, n_{B_N}\bigr)$ 
by
\begin{align}
n_{A_i} &= \max\bigl\{2(N-1)-h_{x_i},0\bigr\} \, ,\\
n_{B_i} &= 1 + n_{A_i}- n_{A_{i-1}} \, . 
\end{align}
for $i=1,\ldots, n$. This defines a reachable state of the SPN, 
since $n_{A_i}\in \{0,\ldots, 2(N_1)\}$ for all $i$, and
$\sum_{i=1}^N n_{B_i}=N$, see Lemmas~\ref{lemma1} and~\ref{Sum_N_b}. 

Observe that the values of $\bigl(n_{B_1},\ldots, n_{B_N}\bigr)$
agree with the previously defined
mapping $\Phi(\eta,x_0)$, provided that
$n_{A_i}>0$ for all $i$. To see this 
we compute from the definition of $\Psi$
\begin{align*}
n_{B_i} \ = \ 1+n_{A_i} -n_{A_{i-1}} \ = \ 
1-\bigl(h_{x_i} -h_{x_{i-1}}\bigr) \ = \ x_i-x_{i-1}-1 \, , 
\end{align*}
where the first equality
is the definition of $\Psi$,
the second holds since $n_{A_i}>0$
for all $i$, and the last uses that the
occupied positions between $x_{i-1}$ and
$x_i$ each contribute $-1$ to
$h_{x_i}-h_{x_{i-1}}$, while the vacant site
at $x_i$ contributes $+1$.
Similar to $\Phi$, the mapping $\Psi$ 
is surjective but not injective.

\begin{proposition}\label{prop:Psi}
Let $h(t)$ be the height function of a TASEP
with $N$ particles on a ring of length $2N$, and let $x_0(t)$ be a 
`tagged' vacant position. Assume the initial state is
\begin{align*}
h_x(0)= \begin{cases} 0, \quad &\text{if $x$ is even,}\\
-1, \quad &\text{if $x$ is odd,} 
\end{cases}
\end{align*}
and set $x_0(0)=2N$.
Then the process $\Psi (h(t),x_0(t))$ agrees with
the Ring AllReduce SPN from Sec.~\ref{sec:spn-model}.
\end{proposition}

\begin{proof} Let $\eta(t)$ be the occupation vector associated 
with $h(t)$, and let $x_i(t)$ the $i$-th vacant 
position clockwise from $x_0(t)$.  The proof proceeds in three steps.
\begin{enumerate}

\item {\em Initial and absorbing states.} 
By definition of $\Psi$, the initial markings
are $n_{A_i}(0)=2(N-1)$ and $n_{B_i}(0)=1$
for $i=1,\ldots, N$. 
As $t\to\infty$, we have that
$h_x(t)\to\infty$,
and $\Psi(h(t), x_0(t))$ reaches the 
absorbing state of the SPN with markings $n_{A_i}=0$, $n_{B_i}=1$ 
for $i=1,\ldots, N$.

\item {\em Evolution prior to wind-down.} 
For $0<t<T_{\rm wind-down}$, we have
$0<n_{A_i}(t) = 2(N-1)-h_{x_i(t)}(t)$, and
$n_{B_i}(t) =x_i(t)-x_{i-1}(t)-1$.
By Eqs.~\eqref{eq:hop-xi} and \eqref{eq:hop-hi}, 
a hop with $y(t^-)=x_i(t^-)-1$ is mapped by $\Psi$ to a 
firing of transition $T_i$ that decrements $n_{A_i}$ and $n_{B_i}$,  and 
increments $n_{B_{i+1}}$, see Eq.~\eqref{eq:MC-update}. 

\item {\em Wind-down}. We finally consider the
phase where
some, but not all workers have completed their $2(N-1)$ firings.
Define the set 
$$
D(t)=\bigl\{i\in \{1,\ldots, N\} \ \big\vert\ h_{x_i(t)}(t)\ge 2(N-1)\bigr\}.
$$
We have already shown that $\Psi(h(t),x_0(t))$
obeys the SPN so long as $D(t)$ is empty, 
i.e., so long as all workers have work remaining.
It suffices to show that in every TASEP state, the transitions
of $\Psi(h(t),x_0(t))$ driven by the TASEP process
agree with the transitions of the SPN.

Consider a hop as in Eq.~\eqref{eq:hop} with $y(t^-)=x_{i}(t^-)-1$.
and assume that $t$ is the first time where the
claim fails.  If $i\in D(t^-)$, then
the TASEP update proceeds as in
Eq.~\eqref{eq:hop-hi}, and therefore $i\in D(t)$.
In the SPN, transition $T_i$ is not enabled 
since $n_{A_i(t)^-}=0$. Therefore $n_{A_i}(t)=0$,
and the markings of $B_i$ and $B_{i+1}$ remain
unchanged.

If $i\not\in D(t^-)$ then $n_{A_i}(t^-)>0$ and the hop 
triggers a firing of transition $T_i$,
in agreement with the SPN. If $h_{x_i(t^-)}(t^-)<2(N-1)-1$
then the update proceeds in the same way as it 
would have prior to wind-down, and $D(t)=D(t^-)$.
If $h_{x_i(t^-)}(t^-)=2(N-1)-1$, then 
in addition to the other updates~$i$ is added to the set $D(t)$,
and $t=T_{{\rm RAR},i}$.
In any case the values of $\max\{2(N-1)-h_{x_j(t)}(t),0\}$ 
and the markings $n_{A_j}(t)$ are not affected for $j\ne i$.
\end{enumerate}
We have shown that a TASEP hop of a particle to
a vacant position at $x_i$ is mapped by
$\Psi$ to a firing of the SPN transition $T_i$
if and only if $T_i$ is enabled in the SPN.
Since all transitions occur at the same exponential rate,
and there are no other transitions, this completes the proof.
\end{proof}

Proposition~\ref{prop:Psi} implies that
the completion time of an individual worker
can be expressed as 
\begin{align}
\label{eq:T-rar-hi}
T_{{\rm RAR},i} &\sim \inf\bigl\{t>0 \ \big\vert\ 
h_{x_i(t)}(t) \ge 2(N-1)\bigr\}\,.
\end{align}

\subsection{TASEP predictions for large rings}
\label{subsec-predictions}

In this subsection, we use recent results
on TASEP to analyze the completion time
$T_{\rm RAR}$ of the Ring AllReduce SPN.
Most of the exact results on TASEP
in the literature concern the long-time asymptotics 
of the height function for TASEP on the line. 
For the {\em `flat'} condition on the height function
\begin{align}
\label{eq:initial-TASEP-line}
h_x(0)= \begin{cases} 0, \quad &\text{if $x$ is an even integer,}\\
-1, \quad &\text{if $x$ is an odd integer,}
\end{cases}
\end{align}
it is known that
\begin{align}
\label{eq:TASEP-asymp}
h_x(t) \sim Ct - t^{1/3} \zeta\qquad\text{as}\ t\to\infty\,.
\end{align}
Here $C$ is a constant, $\zeta$ a random variable,
and the symbol $\sim$ signifies
that $t^{-1/3}\bigl(Ct-h_x(t)\bigr)\to \zeta$ in distribution.
The distribution of $\zeta$ is a known
special function, which we will not discuss here.

The asymptotic in Eq.~\eqref{eq:TASEP-asymp}
remains valid for a ring
of length $2N$ in the limit where both $t,N\to\infty$, 
so long as $t\ll N^{3/2}$.  
The reason is that at any time $t>0$,
the values of the height function at two points $x,y$ are 
almost independent if $|x-y|\gg t^{2/3}$,
but have significant positive correlation
if $|x-y|\ll t^{2/3}$. Consequently
the ring structure does not 
manifest itself in the dynamics until long after the
expected completion time $T^\ast_{\rm RAR}
= 2(N-1)/(1-\mathsf{P}^\ast_{\rm idle})$ \cite{Liu26tasep,Sly2026tasep}.

We next obtain an asymptotic 
formula corresponding to Eq.~\eqref{eq:TASEP-asymp}
for the distribution of $T_{{\rm RAR},i}$.
Fix $i\in \{1,\ldots, N\}$.
For $r\ge 0$ let $\tau_{i, r}$ be the 
time when Worker$_i$ fires for the $r$-th time,
as defined in Eq.~\eqref{eq:tau-ir}.
By Proposition~\ref{prop:Psi},
\[ \tau_{i,r}=\inf\{t>0\mid h_{x_i(t)}(t)\ge r\}
\]
for $0\le r\le 2(N-1)$.  Observe that
$$\tau_{i,r}>t \quad\Longleftrightarrow\quad h_{x_i(t)}(t)<r\,,$$
since both relations describe the event
that Worker$_i$ has completed fewer than $r$ firings by time $t$.
The probability of these events
does not depend on $i$. In fact, $h_{x_i(t)}(t)$ has
the same distribution as $h_x(t)$ 
for an arbitrary vacant position, $x$.
Setting $t=(n + n^{1/3}s)/C$ in Eq.~\eqref{eq:TASEP-asymp},
it follows that 
\begin{align*}
Pr\bigl(\tau_{i,r}>r + r^{1/3}s\bigr)
& = Pr\bigl(h_x(r+ r^{1/3}s) < r\bigr)\\
& = Pr\bigl(C(r + r^{1/3}s)- \bigl(r+r^{1/3}s\bigr)^{1/3}
\zeta  <r\bigr)\\
& = Pr\left(\zeta> \left(\frac{Csr}{r+r^{1/3}s}\right)^{1/3}\right)
\quad \longrightarrow \quad Pr\left(\zeta>Cs \right) \, , 
\end{align*}
as $r, N\to\infty$ with $r\ll N^{3/2}$.  
For $r=2(N-1)$, we match the linear term
to the formula for $T^\ast_{\rm RAR}$ in Eq.~\eqref{eq:Tss},
and set $C=\lim_{N\to\infty}(1-\mathsf{P}^\ast_{\rm idle})=\frac12$.
This establishes Eq.~\eqref{eq:TASEP-i} with $\xi = 2^{4/3}\zeta$.

\bigskip
We turn to the completion time $T_{\rm RAR}= \max_i T_{{\rm RAR},i}$.
Using the TASEP scaling described above, we
write 
$$T_{RAR}=\mathbb{E}\left[T^\ast_{RAR}\right] + 
N^{1/3} \max_i \xi_i\,,
$$
where $\xi_i$ are the random variables defined in Eq.~\eqref{eq:xi-i}. 
for $i=1,\ldots, N$.
If the $\xi_i$
were independent and exponentially
distributed, Extreme Value Theory would
predict that the expected value of the maximum
grows proportionally to $\log N$, while
the standard deviation remains constant.
For distributions with faster decaying tails, 
e.g., a Gaussian, the value of the maximum grows
with a smaller power of $\log N$, and
the standard deviation decays with $\log N$.

Since the random variables $\xi_i$ are dependent, we cannot
directly apply Extreme Value Theory, and therefore resort
to a heuristic analysis.
Using that $\xi_i$ and $\xi_j$ are almost
independent if $|i-j|$ is large compared to $N^{2/3}$,
we approximate $\max_{i=1,\ldots, N}\xi_i$
by the maximum of approximately $N^{1/3}$ i.i.d random
variables.  We observe that the curve for $\mathbb{E}\bigl[T_{\rm RAR}\bigr]$ 
in Fig.~\ref{fig:wind_down_scaled}
indeed grows roughly linearly with $\log N$, 
yielding Eq.~\eqref{eq:TASEP-RAR}.
A corresponding analysis of 
$$T_{RAR}=\mathbb{E}\left[T^\ast_{RAR}\right] + 
N^{1/3} \min_i \xi_i\,,
$$
predicts a linear decay of the curve
for $T_{\rm wind-down}$ in Fig.~\ref{fig:wind_down_scaled},
as given in Eq.~\eqref{eq:TASEP-wd}.

%% file: 6_Conclusion.tex
\section{Conclusions}
\label{sec:conclusion}
We presented a stochastic performance evaluation of the Ring AllReduce
collective communication primitive where we analyzed 
completion times and per-worker idle times. 
We developed an SPN model to represent the dependency
structure among workers in Ring AllReduce and derived 
analytical results for idle probability and completion-time behavior 
across the ramp-up, stable, and wind-down phases.
In the homogeneous setting, we derived the steady-state idle probability of 
an SPN with exponential and shifted exponential firing delays. 
For a heterogeneous setting with stragglers, we 
conducted a mean value analysis of a product form queueing network. 
By coupling the SPN model of Ring AllReduce to a TASEP,  
we were able to characterize the distribution of completion times. 
The two main findings of this paper are that (1) statistical fluctuations 
during Ring AllReduce leaves workers substantially underutilized, 
and (2) due to the tight coupling of operations in a logical ring, 
performance metrics in Ring AllReduce exhibit anomalous scaling. 
Since Ring AllReduce is a building block of many variants of the AllReduce 
operation, e.g., 2D Ring, 2D Torus, Multi-Ring, the 
analysis in this paper can be used to evaluate trade-offs of 
other collective communication primitives.

%% file: RingAllreduce-arxiv-v1.bbl
\begin{thebibliography}{10}
\providecommand{\url}[1]{#1}
\csname url@samestyle\endcsname
\providecommand{\newblock}{\relax}
\providecommand{\bibinfo}[2]{#2}
\providecommand{\BIBentrySTDinterwordspacing}{\spaceskip=0pt\relax}
\providecommand{\BIBentryALTinterwordstretchfactor}{4}
\providecommand{\BIBentryALTinterwordspacing}{\spaceskip=\fontdimen2\font plus
\BIBentryALTinterwordstretchfactor\fontdimen3\font minus
  \fontdimen4\font\relax}
\providecommand{\BIBforeignlanguage}[2]{{%
\expandafter\ifx\csname l@#1\endcsname\relax
\typeout{** WARNING: IEEEtran.bst: No hyphenation pattern has been}%
\typeout{** loaded for the language `#1'. Using the pattern for}%
\typeout{** the default language instead.}%
\else
\language=\csname l@#1\endcsname
\fi
#2}}
\providecommand{\BIBdecl}{\relax}
\BIBdecl

\bibitem{jeon2019analysis}
M.~Jeon, S.~Venkataraman, A.~Phanishayee, J.~Qian, W.~Xiao, and F.~Yang,
  ``Analysis of large-scale multi-tenant {GPU} clusters for {DNN} training
  workloads,'' in \emph{Proc. USENIX Annual Technical Conference (USENIX ATC
  19)}, 2019, pp. 947--960.

\bibitem{hu2024characterization}
Q.~Hu \emph{et~al.}, ``Characterization of large language model development in
  the datacenter,'' in \emph{Proc. 21st USENIX Symposium on Networked Systems
  Design and Implementation (NSDI 24)}, 2024, pp. 709--729.

\bibitem{jiang2024megascale}
Z.~Jiang \emph{et~al.}, ``{MegaScale}: Scaling large language model training to
  more than 10,000 {GPUs},'' in \emph{21st USENIX Symposium on Networked
  Systems Design and Implementation (NSDI 24)}, 2024, pp. 745--760.

\bibitem{Grattafiori2024}
A.~Grattafiori \emph{et~al.}, ``The {Llama} 3 herd of models,'' \emph{arXiv
  Preprint arXiv:2407.21783}, 2024.

\bibitem{AlqahtaniDemirbas2019}
S.~Alqahtani and M.~Demirbas, ``Performance analysis and comparison of
  distributed machine learning systems,'' \emph{arXiv Preprint
  arXiv:1909.02061}, 2019.

\bibitem{Patarasuk2009}
P.~Patarasuk and X.~Yuan, ``Bandwidth optimal all-reduce algorithms for
  clusters of workstations,'' \emph{Journal of Parallel and Distributed
  Computing}, vol.~69, no.~2, pp. 117--124, 2009.

\bibitem{ringallreduce-uber}
A.~Sergeev and M.~Del~Balso, ``Horovod: fast and easy distributed deep learning
  in tensorflow,'' \emph{arXiv Preprint arXiv:1802.05799}, 2018.

\bibitem{spitzer1970}
F.~Spitzer, ``Interaction of {M}arkov processes,'' \emph{Advances in
  Mathematics}, vol.~5, no.~2, pp. 246--290, 1970.

\bibitem{quastel2007}
J.~Quastel and B.~Valk{\'o}, ``$t^{1/3}$ superdiffusivity of finite-range
  asymmetric exclusion processes on,'' \emph{Communications in Mathematical
  Physics}, vol. 273, no.~2, pp. 379--394, 2007.

\bibitem{ringallreduce-baidu}
\BIBentryALTinterwordspacing
A.~Gibiansky, ``Bringing {HPC} techniques to deep learning,'' Baidu Research,
  Technical Report, 2017. [Online]. Available:
  \url{https://andrew.gibiansky.com/blog/machine-learning/baidu-allreduce}
\BIBentrySTDinterwordspacing

\bibitem{Hockney1994}
R.~W. Hockney, ``The communication challenge for {MPP}: {Intel Paragon} and
  {Meiko CS-2},'' \emph{Parallel Computing}, vol.~20, no.~3, pp. 389--398,
  1994.

\bibitem{Thakur2005}
R.~Thakur, R.~Rabenseifner, and W.~Gropp, ``Optimization of collective
  communication operations in {MPICH},'' \emph{International Journal of High
  Performance Computing Applications}, vol.~19, no.~1, pp. 49--66, 2005.

\bibitem{Chan2007}
E.~Chan, M.~Heimlich, A.~Purkayastha, and R.~van~de Geijn, ``Collective
  communication: {T}heory, {P}ractice, and {E}xperience,'' \emph{Concurrency
  and Computation: Practice and Experience}, vol.~19, no.~13, pp. 1749--1783,
  2007.

\bibitem{Barrachina2023}
S.~Barrachina, A.~Castell{\'o}, M.~Catal{\'a}n, M.~F. Dolz, and J.~I. Mestre,
  ``Using machine learning to model the training scalability of convolutional
  neural networks on clusters of {GPUs},'' \emph{Computing}, vol. 105, pp.
  915--934, 2023.

\bibitem{Naumov2020}
M.~Naumov \emph{et~al.}, ``Deep learning training in {Facebook} data centers:
  Design of scale-up and scale-out systems,'' \emph{arXiv Preprint
  arXiv:2003.09518}, 2020.

\bibitem{Nuriyev2021}
E.~Nuriyev and A.~L. Lastovetsky, ``Efficient and accurate selection of optimal
  collective communication algorithms using analytical performance modeling,''
  \emph{IEEE Access}, vol.~9, pp. 109\,355--109\,373, 2021.

\bibitem{Nuriyev2024SUARA}
E.~Nuriyev, R.~R. Manumachu, S.~Aseeri, M.~K. Verma, and A.~L. Lastovetsky,
  ``{SUARA}: A scalable universal allreduce communication algorithm for
  acceleration of parallel deep learning applications,'' \emph{Journal of
  Parallel and Distributed Computing}, vol. 183, p. 104767, 2024.

\bibitem{alqahtani2019performance}
S.~Alqahtani and M.~Demirbas, ``Performance analysis and comparison of
  distributed machine learning systems,'' in \emph{Proc. 28th International
  Conference on Computer Communications and Networks (ICCCN)}.\hskip 1em plus
  0.5em minus 0.4em\relax IEEE, 2019, pp. 1--11.

\bibitem{tensorflow2015-whitepaper}
\BIBentryALTinterwordspacing
M.~Abadi \emph{et~al.}, ``{TensorFlow}: Large-scale machine learning on
  heterogeneous systems,'' 2015, software available from tensorflow.org.
  [Online]. Available:
  \url{https://www.tensorflow.org/extras/tensorflow-whitepaper2015.pdf}
\BIBentrySTDinterwordspacing

\bibitem{li2024straggler}
Y.~Li \emph{et~al.}, ``Straggler-aware gradient aggregation for large-scale
  distributed deep learning system,'' \emph{IEEE/ACM Transactions on
  Networking}, vol.~32, no.~6, pp. 4917--4930, 2024.

\bibitem{chen2024rina}
Z.~Chen, X.~Liu, M.~Li, Y.~Hu, H.~Mei, H.~Xing, H.~Wang, W.~Shi, S.~Liu, and
  Y.~Xu, ``Rina: Enhancing ring-allreduce with in-network aggregation in
  distributed model training,'' in \emph{Proc. IEEE 32nd International
  Conference on Network Protocols (ICNP)}.\hskip 1em plus 0.5em minus
  0.4em\relax IEEE, 2024.

\bibitem{SDRRDMA2028}
M.~Khalilov \emph{et~al.}, ``{SDR-RDMA}: Software-defined reliability
  architecture for planetary scale {RDMA} communication,'' \emph{arXiv Preprint
  arXiv:2505.05366}, 2025.

\bibitem{yu2022scheduling}
M.~Yu, B.~Ji, H.~Rajan, and J.~Liu, ``On scheduling ring-all-reduce learning
  jobs in multi-tenant {GPU} clusters with communication contention,'' in
  \emph{Proc. 23rd International Symposium on Theory, Algorithmic Foundations,
  and Protocol Design for Mobile Networks and Mobile Computing}, 2022, pp.
  21--30.

\bibitem{Yu2022GADGET}
M.~Yu, Y.~Tian, B.~Ji, C.~Wu, H.~Rajan, and J.~Liu, ``{GADGET}: Online resource
  optimization for scheduling ring-all-reduce learning jobs,'' in \emph{Proc.
  IEEE INFOCOM 2022}, 2022, pp. 1569--1578.

\bibitem{zhang2026straggler}
Z.~Zhang and H.~Shen, ``Straggler tolerant and resilient {DL} training on
  homogeneous {GPUs},'' in \emph{2026 35th International Conference on Computer
  Communications and Networks (ICCCN)}, 2026.

\bibitem{Copik2023FMI}
M.~Copik, R.~B{\"o}hringer, A.~Calotoiu, and T.~Hoefler, ``{FMI}: Fast and
  cheap message passing for serverless functions,'' in \emph{Proc. ACM
  International Conference on Supercomputing ({ICS})}, 2023.

\bibitem{barrak2023spirt}
A.~Barrak, F.~Petrillo, and F.~Jaafar, ``{SPIRT}: A fault-tolerant and reliable
  peer-to-peer serverless {ML} training architecture,'' in \emph{Proc. IEEE
  International Conference on Software Quality, Reliability and Security
  ({QRS})}, 2023.

\bibitem{GimenoSarroca2024MLLess}
P.~Gimeno~Sarroca and M.~S{\'a}nchez-Artigas, ``{MLLess}: Achieving cost
  efficiency in serverless machine learning training,'' \emph{Journal of
  Parallel and Distributed Computing}, vol. 183, p. 104764, 2024.

\bibitem{besozzi2026hpcserverless}
V.~Besozzi, M.~Della~Bartola, P.~Dazzi, and M.~Danelutto, ``High-performance
  serverless computing: A systematic literature review on serverless for {HPC},
  {AI}, and big data,'' \emph{IEEE Access}, vol.~13, pp. 195\,611--195\,656,
  2025.

\bibitem{peterson1981petri}
J.~L. Peterson, \emph{Petri Net Theory and the Modeling of Systems}.\hskip 1em
  plus 0.5em minus 0.4em\relax Englewood Cliffs, NJ: Prentice-Hall, 1981.

\bibitem{Marsan1995SPN}
M.~A. Marsan, G.~Balbo, G.~Conte, S.~Donatelli, and G.~Franceschinis,
  \emph{Modelling with Generalized Stochastic Petri Nets}, ser. Wiley Series in
  Parallel Computing.\hskip 1em plus 0.5em minus 0.4em\relax Chichester, UK:
  John Wiley \& Sons, 1995.

\bibitem{White1997MSER}
J.~K.~Preston~White, ``An effective truncation heuristic for bias reduction in
  simulation output,'' \emph{Simulation}, vol.~69, no.~6, pp. 323--334, 1997.

\bibitem{WhiteCobbSpratt2000}
J.~K.~Preston~White, M.~J. Cobb, and S.~C. Spratt, ``A comparison of five
  steady-state truncation heuristics for simulation,'' in \emph{Proc. 2000
  Winter Simulation Conference}, 2000, pp. 755--760.

\bibitem{flajolet2009analytic}
P.~Flajolet and R.~Sedgewick, \emph{Analytic Combinatorics}.\hskip 1em plus
  0.5em minus 0.4em\relax Cambridge University Press, 2009.

\bibitem{graham1994concrete}
R.~L. Graham, D.~E. Knuth, and O.~Patashnik, \emph{Concrete Mathematics: A
  Foundation for Computer Science}, 2nd~ed.\hskip 1em plus 0.5em minus
  0.4em\relax Reading, Massachusetts: Addison-Wesley, 1994.

\bibitem{Little2011Reprint}
J.~D.~C. Little, ``Reprint: {Little's} law as viewed on its 50th anniversary,''
  \emph{Operations Research}, vol.~59, no.~3, pp. 536--549, 2011.

\bibitem{gordon1967closed}
W.~J. Gordon and G.~F. Newell, ``Closed queuing systems with exponential
  servers,'' \emph{Operations Research}, vol.~15, no.~2, pp. 254--265, 1967.

\bibitem{bcmp}
F.~Baskett, K.~M. Chandy, R.~R. Muntz, and F.~G. Palacios, ``Open, closed, and
  mixed networks of queues with different classes of customers,'' \emph{Journal
  of the ACM (JACM)}, vol.~22, no.~2, pp. 248--260, 1975.

\bibitem{reiser1980mean}
M.~Reiser and S.~S. Lavenberg, ``Mean-value analysis of closed multichain
  queuing networks,'' \emph{Journal of the ACM}, vol.~27, no.~2, pp. 313--322,
  1980.

\bibitem{TASEP1}
T.~M. Liggett, \emph{Interacting Particle Systems}, ser. Grundlehren der
  Mathematischen Wissenschaften.\hskip 1em plus 0.5em minus 0.4em\relax New
  York: Springer-Verlag, 1985, vol. 276.

\bibitem{TASEP2}
B.~Derrida, ``An exactly soluble non-equilibrium system: The asymmetric simple
  exclusion process,'' \emph{Physics Reports}, vol. 301, no. 1--3, pp. 65--83,
  1998.

\bibitem{tracy-Widom2008tasep}
C.~A. Tracy and H.~Widom, ``Integral formulas for the asymmetric simple
  exclusion process,'' \emph{Communications in Mathematical Physics}, vol. 279,
  no.~3, pp. 815--844, 2008.

\bibitem{BFPS_2007tasep}
A.~Borodin, P.~L. Ferrari, M.~Pr{\"a}hofer, and T.~Sasamoto, ``Fluctuation
  properties of the {TASEP} with periodic initial configuration,''
  \emph{Journal of Statistical Physics}, vol. 129, no.~5, pp. 1055--1080, 2007.

\bibitem{Quastel26}
\BIBentryALTinterwordspacing
J.~Quastel, ``The {KPZ} fixed point,'' in \emph{Proc. International Congress of
  Mathematicians}, S.~Friedlander and Y.~Tschinkel, Eds., vol. 2: Plenary
  Lectures.\hskip 1em plus 0.5em minus 0.4em\relax Society for Industrial and
  Applied Mathematics (SIAM), 2026, pp. 272--301. [Online]. Available:
  \url{https://epubs.siam.org/doi/full/10.1137/25M1805643}
\BIBentrySTDinterwordspacing

\bibitem{Liu26tasep}
J.~Baik and Z.~Liu, ``Tasep on a ring in sub-relaxation time scale,''
  \emph{Journal of Statistical Physics}, vol. 165, no.~6, pp. 1051--1085, 2016.

\bibitem{Sly2026tasep}
D.~Schmid and A.~Sly, ``Mixing times for the tasep on the circle: D. schmid, a.
  sly,'' \emph{Probability Theory and Related Fields}, vol. 194, no.~3, pp.
  1161--1233, 2026.

\end{thebibliography}
